\documentclass[a4paper]{article}
\usepackage{amsmath, amssymb}
\usepackage{a4wide}
\usepackage{graphicx}
\usepackage{hyperref}

\newcommand{\dis}{\displaystyle }
\newcommand{\al}{\alpha }

\newcommand{\pr}{\prime }
\newcommand{\la }{\lambda }

\newcommand{\rar}{\rightarrow}
\newcommand{\be}{\begin{equation}}

\newcommand{\ee}{\end{equation}}
\newcommand{\wt}{\widetilde}
\newcommand{\bearr}{\begin{eqnarray}}
\newcommand{\bearrs}{\begin{eqnarray*}}
\newcommand{\eearr}{\end{eqnarray}}
\newcommand{\eearrs}{\end{eqnarray*}}
\newcommand{\barr}{\begin{array}}
\newcommand{\earr}{\end{array}}
\newcommand{\p}{\partial}

\def\l{\left}
\def\r{\right}
\newcommand{\un}{\underline}

\newcommand{\f}{\frac}

\begin{document}

\title{Cosmology for Particle Physicists}

\author{U. A. Yajnik\\
\textsl{Indian Institute of Technology, Bombay, Mumbai}}
\date{}
\maketitle
\vfill
\thispagestyle{empty}
\begin{center}
Lecture Notes\\
XXI SERC School on Theoretical HEP\\
PRL, Ahmedabad, February 2006\\
\end{center}
\newpage

\tableofcontents
\newpage

\section{Introduction}
\label{sec:introduction}

Over the past two decades Cosmology has increasingly become a precision
science. That the Universe is expanding was an astonishing discovery.
Now we know its details to unprecedented precision. An expanding Universe
also implied an extremely compact state in the past, and therefore very 
high temperature. The Particle Physics forces which can now be explored only
in accelerator laboratories were in free play in the remote past. Thus
the observation of the oldest remnants in the Universe amounts to looking 
at the results of a Particle Physics experiment under natural conditions.

In these notes we present a selection of topics, each section approximately
amounting to one lecture. We begin with a brief
recapitulation of General Relativity, and the Standard Model of Cosmology. 
The study of Cosmology requires General Relativity to be applied only 
under a highly  symmetric situation and  therefore it is possible 
to recast the essentials as Three Laws of Cosmology. The study of very 
early Universe brings us squarely into the domain of Quantized Field 
Theory at given temperature. 
Intermediate metastable  phases through which the Universe passed require 
an understanding of the effective potential of the field theory in a 
thermal equilibrium. This formalism is developed in some detail.

The remainder of the notes discuss important signatures of the remote past. 
These include : (i) inflation, (ii) density perturbations leading to 
galaxy formation, (iii) study of hot and cold relics  decoupled from 
the remaining constituents, some of which can be candidates for Dark Matter, 
(iv) finally the baryon asymmetry of the Universe. As we shall see each 
of these has a strong bearing on Particle Physics and is being subjected 
to ever more precise observations.

\subsection{ General Theory of Relativity : A Recap}\label{sec:GRrecap}

Special Theory of Relativity captures the kinematics of space-time observations.
On the other hand, General Theory of Relativity is a dynamical theory, which
extends the Newtonian law of gravity to make it consistent with Special
Relativity. In this sense it is not a ``generalization'' of Relativity but rather,
a theory of Gravity on par with  Maxwell's theory of Electromagnetism. It is 
nevertheless a very special kind of theory because of the Principle of
Equivalence. The equivalence of gravitational and inertial masses ensures that
in a given gravitational field, all test particles would follow a trajectory 
decided only by their initial velocities, regardless of their mass.
This makes it possible to think of the resulting trajectory as a property 
of the spacetime itself. This was the motivation for introducing methods 
of Differential Geometry, the mathematics of curved spaces for the 
description of Gravity. Due to this ``grand unification'' of space and time
into a dynamically varying set we shall use the convention of writing the word 
spacetime without hyphenation as adopted in ref. [1].

Throughout these notes we use the convention $\hbar=c=1$ and the sign
convention $ds^2=dt^2 - \l| d\mathbf{x}\r|^2$ for the spacetime interval in
Special Relativity. 
The principle of General Covariance states that a given gravitational field
is described by a metric tensor $g_{\mu\nu}$ a function of spacetime
variables collectively written as $x^\mu \equiv \mathbf{x},t$. Gravity
modifies the spacetime interval to the general quadratic form
$g_{\mu\nu}dx^\mu dx^\nu$, where the summation convention on same indices 
is assumed. The trajectories of test particles are simply the shortest 
possible paths in this spacetime, determined by the metric tensor through 
the geodesic  equation
\[
\f{d^2x^\mu}{d\tau^2} + \Gamma^{\mu}_{\nu\rho}\f{dx^\nu}{d\tau} 
\f{dx^\rho}{d\tau} = 0
\]
where the Christoffel symbols $\Gamma^\mu_{\nu\rho}$ are given by
\[
\Gamma^\mu_{\nu\rho} = \f{1}{2}g^{\mu\la}\l( \f{\p g_{\nu\la}}{\p x^\rho} 
- \f{\p g_{\nu\rho}}{\p x^\la} +\f{\p g_{\rho\la}}{\p x^\nu} \r)
\]
These symbols are not tensors but determine the covariant derivative much
the same way that the electromagnetic potentials which are themselves not
gauge invariant determine the minimal coupling of charged particles to
electromagnetic fields.

The equations which determine the gravitational field, i.e., the 
tensor $g_{\mu\nu}$ itself are the Einstein Equations, 
\[
G_{\mu\nu}- \Lambda g_{\mu\nu} \equiv R_{\mu\nu} -\f{1}{2}g_{\mu\nu}R 
- \Lambda g_{\mu\nu} = 8\pi GT_{\mu\nu}
\]
where $T_{\mu\nu}$ is the energy momentum tensor and the Ricci tensor
$R_{\mu\nu}$ and the scalar curvature $R$ are the contracted forms
of the fourth rank tensor the Riemann curvature, given by
\bearrs
R^\mu_{\nu \alpha \beta} &=& \p_\al \Gamma^\mu_{\nu\beta}
- \p_\beta \Gamma^\mu_{\nu\alpha} 
+ \Gamma^\mu_{\sigma\al} \Gamma^\sigma_{\nu\beta}
- \Gamma^\mu_{\sigma\beta} \Gamma^\sigma_{\nu\al} \\[10pt]
R_{\mu\nu}\phantom{\alpha} &=& R^\la_{\mu\la\nu} \\[10pt]
R\phantom{\mu\nu\alpha} &=& g^{\mu\nu}R_{\mu\nu}
\eearrs
The tensor $G_{\mu\nu}$ is called the Einstein tensor and has the elegant
property that its covariant derivative vanishes. The last term on the
left hand side is called the Cosmological term since its effects are not
seen on any small scale, even galactic scales. It can be consistently 
introduced into the equations provided $\Lambda$ is a constant. Since
the covariant derivative $D_\rho g_{\mu\nu}=0$, the covariant derivative 
of this term also vanishes. This fact is matched on the right hand side 
of Einstein's equation by the vanishing of the
covariant derivative of the energy-momentum tensor. The vanishing of 
the Einstein tensor follows from the Bianchi identities in Differential 
Geometry. The geometric significance of the identities is that given a 
spacetime domain, they express the statement``the boundary of a boundary 
is zero''. We leave it to the reader 
to pursue ref [1] to understand its details. Thus a geometric principle
implies the covariant conservation of energy-momentum tensor, a physical
quantity. But it has to be noted that covariant conservation does not 
imply a conserved charge the way it happens in flat spacetime with 
divergence of a four-vector. But if there are $3$-dimensional regions 
on whose  $2$-dimensional boundaries gravity is very weak, it does imply 
conservation of total mass-energy in the given  volume.

There are quite a few subtleties concerning the implications of General
Relativity and the conditions under which it supercedes Newtonian gravity. 
We present here a few ``True or False'' statements for the reader to think
over and discuss with peers or teachers. Starting points to answers are
given in Appendix, sec. \ref{sec:app01}.

\subsubsection{True or False}
\begin{enumerate}
\item Curved spacetime is a necessity of GR due to the underlying 
Special Relativity principle.
\item The invariance of the equations of physics under arbitrary
reparameterisation of spacetime is the essential new content of GR.
\item The notion of energy density becomes meaningless in GR
\item The notion of total energy becomes meaningless in GR
\item Points where some of the metric coefficients vanish are unphysical
\item Points where some of the metric coefficients diverge are unphysical
\item Points where any components of curvature tensor diverge are unphysical
\item Newtonian gravity is insufficient to describe an expanding Universe
and GR is required.
\end{enumerate}

\subsection{The Standard Model of  Cosmology}
Here we summarise the broadest features of our current understanding
of the Universe based on a vast variety of data and modeling.  We
summarise size, age and contents of the Universe as follows
\begin{itemize}
\item There is a strong indication that the Universe is homogeneous 
and isotropic if we probe it at sufficiently large scales, such as 
the scale of clusters of galaxies. The typical scale size is 10 
Megaparsec (Mpc) and larger. At the scale of several tens of Mpc 
the distribution of galaxies is homogeneous.

It is at present not known whether the Universe is finite 
in size or infinite. A finite Universe would be curved as a three
dimensional manifold. An infinite universe could also show curvature.
At present we do not see any signs of such curvature. 

\item Secondly we believe that the Universe has been expanding 
monotonically for a finite time in the past. This gives a finite
age of about 13.7 billion years to our Universe. 

What was "before" this time is not possible to understand within
the framework of classical General Relativity. But Newton's 
gravitational constant suggests a fundamental mass scale, 
called the Planck scale $M_{Pl}=G^{-1/2}\approx 1.2\times 10^{19}$GeV  
and corresponding  extremely small time scale, $10^{-44}$ sec. 
We expect that Quantum theory of Gravity should take over at 
that scale. Unfortunately that theory has not yet been worked 
out due to insufficient data.

\item Finally it has been possible to map the current contents of 
the Universe to a reasonable accuracy. More about them below.
The contents of the Universe can be divided into three types,
\begin{enumerate}
\item  Radiation and matter as known in Particle Physics

\item Dark Matter,
one or more species of particles, at least one of which is necessarily
non-relativistic and contributing significantly to the total energy
density of the Universe today. These particles have no Standard Model 
interactions. 

\item Dark Energy, the largest contributor to the energy
density balance of the present Universe, a form of energy which does
not seem to fit any known conventional fields or particles. It
could be the discovery of a non-zero cosmological constant. But if
more precise observations show its contribution to be changing with time,
it can be modelled as a relativistic continuum which possesses negative pressure.

\end{enumerate}
\end{itemize}

Of the contents of type 1, there are approximately 400 photons per cc
and $10^{-7}$ protons per cc on the average. Compared to Avogadro
number available on earth, this is a very sparse Universe.
Of these the major contributor to energy density is
baryonic matter. This constitutes stars, galaxies and large Hydrogen
clouds out of which more galaxies and stars are continuously in
formation. The other component is the Cosmic Microwave Background 
Radiation (CMBR), the gas of photons left over after neutral Hydrogen 
first formed.  Its contribution to total energy density is relatively 
insignificant. But its precision study by experiments such as the Wilkinson Microwave
Anisotropy Probe ( WMAP) is providing us with a very detailed information
of how these photons have been created and what are the ingredients
they have encountered on their way from their origin to our probes.

There probably are other exotic forms of energy-matter not yet discovered.
Two principle candidates are topological defects such as cosmic strings,
and the axion. We shall not be able to discuss these here. Axions
are almost a part of Standard Model though very weakly interacting,
and could potentially be Dark Matter candidates.

Aside from these current parameters of the Universe there is a reason
to believe that the Universe passed through one or more critical phases 
before arriving at the vast size and great age it presently has.
This critical phase in its development is called Inflation. It is expected
to have occurred in remote past at extremely high energies, perhaps in
the Planck era itself. What is interesting is that the fluctuations
in energy density which finally became galaxies could have originated 
as quantum effects during that era. Thus we would be staring at the 
results of quantum physics in the very early Universe whenever we 
see galaxies in the sky.

A quantitative summary of present observables of the Universe 
is given at the end of sec. \ref{sec:sec03}.

\subsection{The Standard Model of Particle Physics}

We assume that the reader is familiar with the Standard Model of Particle 
Physics. Appendix B of Kolb and Turner contains a review. Cosmology has a 
dual role to play in our understanding of the  fundamental forces.  
It is presenting us with the need to make further extensions of the 
Standard Model and is also providing
evidence to complete our picture of Particle Physics. For example whether
Dark Matter emerges from an extension of the Standard Model is a challenge
to model building. On the other hand axions expected from QCD may 
perhaps get verified in Astroparticle physics experiments and may have played
a significant role in the history of the Cosmos.

Study of Cosmology also sharpens some of the long recognized problems 
and provides fresh perspectives and fresh challenges. Symmetry breaking
by Higgs mechanism in Standard Model (and its extensions the Grand Unified 
models) causes  hierarchy problem. But it also implies a cosmological constant 
far larger than observed. We hope  that the two problems have a common solution. 
Despite the conceptual problems with the QFT of scalar fields, inflation
is best modeled by a scalar field. Similarly, consider Dark Energy which
is almost like a
cosmological constant of a much smaller value than Particle Physics
scales. This has finally been confirmed over the past decade. Again
many models seem to rely on unusual dynamics of a scalar field to explain
this phenomena. We hope that  supersymmetry or superstring theory will 
provide natural candidates for such  scalar fields without the attendant QFT problems.

\section{ Friedmann-Robertson-Walker Metrics}\label{sec:FRWmetric}

Cosmology began to emerge as a science after the construction of reflection 
telescopes of 100 to 200 inch diameter in the USA at the turn of 1900. 
When Doppler shifts of Hydrogen lines of about twenty nearby galaxies 
could be measured it was observed that they were almost all redshifts. 
Edwin Hubble proposed a linear  law relating redshift and distance. 
Then the data could be understood as a universal expansion.
Over the last 75 years this  fact has been further sharpened, with 
more  than 10 million galaxies observed and cataloged.

It is reasonable to believe that we are not in a particularly special 
galaxy. So it is reasonable to assume that the expansion is uniform, 
i.e. observer on any other galaxy would
also see all galaxies receding from him or her at the same rate. 
This reasoning allows  us to construct a simple solution to Einstein's 
equations. We assume  that the Universe corresponds to a solution 
in which aside from the overall expansion as a function of time, 
the spacetime is homogeneous and isotropic. This would 
of course be true only for the class of observers who are drifting 
along with the galaxies in a systematic expansion. ( An intergalactic 
spaceship moving at a high speed would see the galactic distribution 
quite differently). 
We characterize this coordinate system as one in which (1) there is 
a preferred time coordinate\footnote{This time coordinate is not unique.
We can define a new time coordinate $\tilde t(t)$ as an arbitrary 
smooth function of the old time coordinate $t$. What is unique is
that  a time axis is singled out from the spacetime continuum. What 
units, (including time dependent ones), we use to measure the time 
is arbitrary.} such  that at a given 
instant of this  time, (2) the distribution of galaxies is homogeneous 
and isotropic. 

It should be noted that this is a statement about the symmetry of 
the sought after solution of the Einstein equations. The symmetries 
restrict the boundary conditions under which the equations are 
to be solved and in this way condition the answer. In older literature 
the existence of such symmetries of the observed Universe was 
referred to as Cosmological Principle, asserted in the absence of 
any substantial data. Over the years with accumulation of data we
realize that it is not a principle of physics, but a useful model 
suggested by observations, similar to  the assumption that the 
earth is a spheroid.

The assumptions of homogeneity and isotropy made above are very strong 
and we have a very simple class of metric tensors left as possible
solutions. They can be characterized by spacetime interval of the form
\[
ds^2 = dt^2 - R^2 (t) \l\{ \f{dr^2}{1-kr^2} + r^2d\theta^2 + r^2
\sin^2\theta d\phi^2\r\} \; \;
\]
The only dynamical degree of freedom left is the scale factor $R(t)$.
Further, there are three possibilities distinguished by whether the
spacelike hypersurface at a given time is flat and infinite ( Newtonian
idea), or compact like a ball of dimension $3$ with a given curvature,
(generalization of the $2$ dimensional shell of a sphere in usual
3 dimensional Euclidean space), 
or unbounded and with constant negative curvature, a 
possibility more difficult to visualize. These three possibilities 
correspond to the parameter $k = 0$, or $k=+1$, or $k=-1$. 
The cases $k=-1$ and $k=1$ also have representations which make their 
geometry more explicit 
\bearrs
ds^2 = \l\{ \barr{l} dt^2 - R^2(t) \l\{ d\chi^2 + \sin^2\chi \l(
d\theta^2 + \sin^2\theta d\phi^2\r) \r\}\;\;\; k = 1 \\[10pt]
 dt^2 - R^2(t) \l\{ d\chi^2 + \sinh^2 \chi \l( d\theta^2 +
\sin^2\theta d\phi^2\r) \r\}\;\; k = -1  \earr \right.
\eearrs

The time coordinate $t$ we have used above is a particular choice 
and is called comoving time. An alternative time coordinate $\eta$ 
is given by 
\bearrs &&d\eta= \f{dt}{R(t)}\\[10pt]
ds^2&=& R^2(\eta ) \l\{ d\eta^2 - \f{dr^2}{1-kr^2} -
dr^2 - r^2 \sin^2 \theta d\phi^2 \r\}\\[10pt]
\eearrs
Its advantage is that for $k=0$ it makes the metric conformally
equivalent to flat (Minkowski ) space.

\subsection{Cosmological redshift}
In this and the next subsection we identify the precise definitions
of redshift and cosmological distances to understand Hubble Law in
its general form.

The observed redshift of light is similar to Doppler shift, but we 
would like to think that it arises because spacetime itself is 
changing and not due to relative motion in a conventional sense. 
In other words, in cosmological context the redshift 
should be understood as arising from the fact that time elapses  at a 
different rate at the epoch of emission $t_1$ from that at the epoch $t_0$ 
of  observation.  Since light follows a geodesic obeying $ds^2=0$, it 
is possible for us to define the quantity $f(r_1)$ which is a dimensionless 
measure of the separation between emission point $r=0$ and the observation 
point $r=r_1$
\[
f\l( r_1\r) \equiv {\dis \int}^{r_1}_0 \f{dr}{\l( 1-kr^2\r)^{1/2}} 
= {\dis \int}^{t_0}_{t_1} \f{dt}{R(t)} \\[10pt]
\]
where the second equality uses $ds^2=0$. Now the same $f(r_1)$ is valid
for a light signal emitted a little later at $t=t_1+\delta t_1$ and 
received at a corresponding later time $t_0+\delta t_0$.
\[
f\l( r_1\r) = {\dis \int}^{t_0+\delta t_0}_{t_1 + \delta t_1} \f{dt}{R(t)}
\]
Equivalently,
\bearrs
{\dis \int}^{t_1 + \delta t_1}_{t_1} \f{dt}{R}&=& {\dis
\int}^{t_0+\delta t_0}_{t_0} \f{dt}{R(t)}\\[10pt]
\f{\delta t_1}{R\l( t_1 \r)}&=& \f{\delta t_0}{R \l( t_0 \r)} 
\eearrs
or
\bearrs \f{\nu_0}{\nu_1} =\f{\delta t_1}{\delta t_0} =
\f{\la_1}{\la_0}& = &\f{R\l( t_1 \r)}{R \l( t_0
\r)}\\[10pt]
\eearrs
It is convenient to define the redshift $z$, originally so defined because 
it would always  be small, as given by
\[
1 + z \equiv \f{\la_0}{\la_1}= \f{R\l( t_0 \r)}{R \l( t_1 \r)}
\]

\subsection{ Luminosity Distance}
Defining a measure of spacelike distances is tricky in Cosmology because
physical separations between comoving objects are not static and 
therefore lack an operational meaning. Since distances are measured effectively
by observing light received, we define luminosity distance $d_L$ by
\[
d^2_L = \f{L}{4\pi F} \begin{array}{l}\rar \mbox{absolute luminosity}\\
\rar\mbox{observed flux}
\end{array}
\]
If the metric were frozen to its value at time $t_0$ this would 
have been the same as in flat space, $R(t_0)^2 r_1^2 $ with $r_1$ 
the coordinate distance travelled by light.Due to expansion effects, we 
need additional factors of $1+z$, once for reduction in energy due to 
redshift and once due to delay in the observation of the signal
\[
d^2_L = R^2 \l( t_0 \r) r^2_1 \l( 1 + z \r)^2
\]
We now introduce measures $H_0$ for the first derivative, representing 
the Hubble parameter and  a dimesionless measure $q_0$ for the second 
derivative, traditionally {\it deceleration}, by expanding the scale 
factor as
\bearrs
\f{R(t)}{R\l( t_0 \r)}&=& 1 + H_0 \l( t - t_0 \r) - \f{1}{2} q_0
H^2_0 \l( t - t_0 \r)^2 + ....\\[10pt]
H_0&\equiv& \f{\dot{R}\l( t_0 \r)}{R\l( t_0 \r)} \;\;\;\; q_0
\equiv -\f{\ddot{R} \l( t_0 \r)}{\dot{R}^2 \l( t_0 \r)} R \l( t_0
\r) = \f{-\ddot{R}}{RH^2_0}
 \eearrs
\[
\therefore \; z = H_0 \l( t_0 - t \r) + \l( 1 + \f{q_0}{2} \r)
H^2_0 \l( t_0 - t \r)^2 + ...
\]
so that
\[
\l( t_0 - t \r) = H^{-1}_0 \l( Z - \l( 1 + \f{q_0}{2} \r) Z^2 +
... \r)
\]
We now use the quantity $f(r)$ introduced in the discussion of redshift of light,
which for the three different geometries works out to be
\[
f \l( r_1 \r)= \l\{ \barr{ccc} \sin^{-1} r_1&=& r_1 + \f{\mid
r\mid^3}{6}\\ r_1&&\\ \sinh^{-1}&=& - \f{1}{6} \earr \right.
\]
\[
\therefore r_1 = \f{1}{R \l( t_0 \r)} \l[ \l( t_0 - t_1 \r) +
\f{1}{2} H_0 \l( t_0 - t_1 \r)^2 + ... \r]
\]
Substitute $\l( t_0 - t_1 \r)$
\bearrs
r_1&=& \f{1}{R\l( t_0 \r) H_0} \l[ z - \f{1}{2} \l( 1 +
q_0 \r) z^2 + ... \r]\\[10pt]
H_0d_L&=& z + \f{1}{2} \l( 1 - q_0 \r) z^2 + ... 
\eearrs
The last relation expresses the connection between observed luminosity of
distance $d_L$ of galaxies and their redshift $z$, incorporating the 
curvature effects  arising from the expansion of the Universe. 
Extensive data on $d_L$ and $z$ gathered from galaxy surveys such as
2-degree Field  Galactic Redshift Survey ( 2dF GRS) and 6dF GRS can 
be fitted with this equation to determine cosmological parameters $H_0$
and $q_0$. 

\section{The Three Laws of Cosmology}
\label{sec:sec03}

It is possible to discuss the cosmological solution without recourse 
to the full Einstein equations. After all a comprehensive framework 
like electromagnetism was discovered only as a synthesis of a variety 
of laws applicable to specific situations. Specifically, Coulomb's law 
is most useful for static pointlike charges. Similarly, given that 
the Universe has a highly symmetric distribution of matter, allows 
us to express the physics as three laws applicable to Cosmology, these being 
\begin{enumerate}
\item \un{Evolution of the scale factor} : The evolution of $R(t)$ introduced above
is governed by
\[
\l( \f{\dot{R}}{R}\r)^2 + \f{k}{R^2} -\f{\Lambda}{3} = \f{8\pi}{3} G\rho
\]
\item \un{Generalized thermodynamic relation} : The energy-matter source
of the gravitational field obeys generalization of the first law of 
thermodynamics $dU = -pdV$,
\[
\f{d}{dt} \l( \rho R^3\r) + p \f{d}{dt} \l( R^3 \r) = 0 \; \; \;
\]
To put this to use we need the equation of state $p=p(\rho)$. For most
purposes it boils down to a relation 
\[
p = w\rho
\]
with $w$ a constant.
\item \un{Entropy is conserved} : (except at critical thresholds)
\[
\f{d}{dt} (S) = \f{d}{dt} \l( \f{R^3}{T} (\rho + p)\r) = 0
\]
\end{enumerate}

The law 2 can be used to solve for $\rho$ as a function of $R$ with
$w$ a given parameter. Then this $\rho$ can be substituted in law 1
to solve for $R(t)$ the cosmological scale factor. 

As for Law 3, in the cosmological context we speak of entropy density $s$. 
Thus the above law applies to the combination $sR^3$. Further, non-relativistic
matter does not contribute significantly to entropy while for radiation
$s\propto T^3$. Hence we get the rule of thumb  
\[
S = sR^3 \propto T(t)^3R(t)^3 = constant
\]
Thus Law 3 provides
the relation of the temperature and the scale factor. However, 
critical threshold events are expected to have occured in the
early Universe, due to processes going out of equilibrium or due to 
phase transitions. The constant on the right 
hand side of $R(t) T(t)$ equation has to be then re-set by calculating 
the entropy  produced in such events.

This formulation is sufficient for studying most of Cosmology.
However if we are familiar with Einstein's theory the above
laws can  be derived systematically. We need to calculate
the Einstein tensor. The components of the Ricci tensor and the
scalar curvature $R$ in terms of the scale factor $R(t)$ are
\bearrs 
R_{00}&=& - 3 \f{\ddot{R}}{R}\\[10pt]
R_{ij}&=& - 6 \l( \f{\ddot{R}}{R} + \f{\dot{R}^2}{R^2} +
\f{k}{R^2}
\r) g_{ij}\\[10pt]
R&=& -6 \l( \f{\ddot{R}}{R} + \l( \f{\dot{R}}{R} \r)^2 +
\f{k}{R^2}
\r)\\[10pt]
\eearrs
The resulting Einstein equations contain only two non-trivial
equations, the $G_{00}$ component and the $G_{ii}$ component where
$i=1,2,3$ is any of the space components.
\bearrs 00&:& \l( \f{\dot{R}}{R} \r)^2 + \f{k}{R^2} =
\f{8\pi}{3} G\rho\\[10pt]
ii&:& 2 \f{\ddot{R}}{R} + \l( \f{\dot{R}}{R}\r)^2 + \f{k}{R^2} =
-8\pi Gp 
\eearrs

It turns out that the $ii$ equation can be obtained by differentiating
the $00$ equation and using the thermodynamic relation
\[
d\l( \rho R^3 \r) = -pd\l( R^3 \r)
\]
Hence we only state the second law rather than the $ii$ equation.

\subsection{Example : Friedman Universe }
An early model due to A. A. Friedman (1922) considers a 
$k=1$ universe with pressureless dust, i.e., $p=0$ and with $\Lambda=0$. 
Then according to Law 2,
\[
\f{d}{dt} \l( \rho R^3 \r) = 0
\]
Now let $t_1$ be a reference time so that for any other time $t$, $\rho(t)R^3(t)= \rho(t_1)R^3(t_1)$
Then according to Law 1, 
\bearrs 
\f{1}{R^2} \l( \f{dR}{dt} \r)^2 + \f{k}{R^2} &=&
\f{8\pi}{3} G\rho_1 \f{R_1^3}{R^3}\\[10pt]
\implies \dot{R}^2 - \f{R_{max}}{R} &=& -1  
\eearrs
where
\[
R_{max} \equiv \f{8\pi}{3} G\rho_1 R_1^3 
\]
This equation is first order in time, but non-linear. It can be solved 
by expressing both $t$ and $R$ in terms of another parameter $\eta$. 
The solution is
\[
R(\eta ) = \f{1}{2} R_{max} (1-\cos \eta ) \qquad t(\eta ) = \f{1}{2}
R_{max} (\eta - \sin \eta )
\]
It results in a shape called cycloid.
The first order equation thus solved can also be thought of as a 
particle in potential $V(x) = - \f{x_{max}}{x}$ with
total energy $-1$.\\

\un{Exercise} :  ( Friedman 1924) Solve for the scale factor of a 
$\Lambda=0$ universe with 
pressureless  dust but  with negative constant curvature, i.e., $k=-1$.

\subsection{Parameters of the Universe}
\label{sec:parameters}
We now summarise the observable parameters of the Universe as available from 
several different data sets. But first we introduce some standard conventions.
First we rewrite the first law including the cosmological constant,
\[ 
H^{2}+{k\over R^{2}}-\f{\Lambda}{3} ={8\pi G\over 3}\rho  
\]
Suppose have a way of measuring each of the individual terms in the above equation.
Then with all values plugged in, the left hand side must balance the right hand side.
Our knowledge of the Hubble constant $H_0$ is considerably more accurate than our
knowledge of the average energy density of the Universe. We express the various
contributions in the above equation as fractions of the contribution of the Hubble
term. This can of course be done at any epoch. Thus, dividing out by $H^2$, 
\[ 
1+{k\over H^{2}R^{2}}=\Omega _{\Lambda }+\Omega _{\rho } 
\]
where we define the fractions 
\[
\Omega_\Lambda = \f{\Lambda}{3 H^2}\qquad \Omega_\rho = \f{8\pi G \rho}{3 H^2}
\]
In detail, due to several different identifiable contributions to $\rho$ from
baryons, photons and Dark Matter, we identify individual constributions again as
fractions of the corresponding Hubble term as $\Omega_b$, $\Omega_\gamma$, and 
$\Omega_{DM}$ respectively. Either the sum of the various $\Omega$'s must add
up to unity or the $k\neq 0$. In table \ref{tab:parameters} we list the 
current values of the parameters.

Determination of Hubble constant has several problems of calibration.
It is customary to treat its value as an undetermined factor $h$
times a convenient value $100$ km/s/Mpc which sets the scale of
the expansion rate. It is customary to state many of the parameters 
with factors of $h$ included since they are determined conditionally.
At present $h^2\approx 1/2$. The Dark Energy component seems to behave 
almost like a cosmological constant and hence its contribution is
given the subscript $\Lambda$.
\begin{table}
\begin{center}
\begin{tabular}{ll}
 Age \(t_0 \) & \( 13.7 \pm 0.2\)G. yr,  \\ 
Hubble constant \(H_0\) & \(h \times 100\)km/s/Mpc \\ 
Parameter $h$ & \(  0.71\) or \( 0.73 \) \\
\( T_{CMB}\) & \( 2.725 \times 10^6 \mu\)K  \\
\( \Omega\) & \( 1.02 \pm\  0.02 \) \\
\( \Omega_{\mbox{all matter}}h^2 \)  & \(  0.120 to 0.135   \) \\ 
\( \Omega_bh^2\) & \( 0.022 \) \\
\( \Omega_{\Lambda} \) & \(  0.72 \pm\  0.04,\) \\
\( w\  (\equiv p/\rho )\)  & \( -0.97 \pm 0.08 \)
\label{tab:parameters}
\end{tabular}
\end{center}
\end{table}
$\Omega$ being 
close to unity signifies that we seem to have measured all contributions to energy
density needed to account for the observed Hubble constant. The parameter
$w$ exactly $-1$ corresponds to cosmological constant. Current data are best
fitted with this value of $w$ and assumption of cold, i.e., non-relativistic
Dark Matter. This Friedman model is referred to as the $\Lambda$-CDM model.

\section{The Big Bang Universe}
By its nature as a purely attractive force, gravity does not generically
allow static solutions. 
Since the Universe is isotropic and homogeneous now it is reasonable to
assume that that is the way it has been for as far back as we can extrapolate. 
Such an extrapolation however will require the Universe to have passed through
phases of extremely high densities and pressures, phases where various microscopic
forces become unscreened long range forces, again subject to overall homogeneity 
and isotropy of the medium. This is the essential picture of a Big Bang Universe, 
extrapolated as far back as the Quantum Gravity era about which we can not say 
much at the present  state of knowledge.   However the intervening epochs encode
into the medium important imprints of the possible High Energy forces that may have
been operative. Thus the early Universe is an interesting laboratory for Elementary 
Particle Physics.

\subsection{Thermodynamic Relations}
For a massless single relativistic species,
\bearrs \rho&=& \f{\pi^2}{30} gT^4\;\;\mbox{for bosons}  \hspace{10mm} \ldots \times
\f{7}{8} \; \mbox{for fermions}\\
n&=& \f{\xi(3)}{\pi^2} gT^3\;\;\mbox{for bosons} \;
 \hspace{10mm} \ldots \times \f{3}{4}\; \mbox{for fermions} 
\eearrs
where $g$ is the spin degeneracy factor. In the early Universe,
a particle species with mass much less than the ambient
temperature behaves just like a relativistic 
particle\footnote{Massless vector bosons have one degree of freedom
less than massive vector bosons. So this has to be accounted
for in some way. In the Standard Model of Particle Physics, masses
arise from spontaneous symmetry breaking which disappears at
sufficiently high temperature and the longitudinal vector boson
degree of freedom is recovered as additional scalar modes of 
the Higgs at high temperature.}.
More generally, we introduce an effective degeneracy $g_*$ for
a relativistic gas which is a mixture of various species,
\bearrs 
\rho^{relat}&=& \f{\pi^2}{30} g_* T^4 \equiv 3p^{relat}\\[10pt]
g_*&=& {\dis \sum_i} g_i \l( \f{T_i}{T} \r)^4 + \f{7}{8} {\dis
\sum_j} g_j \l( \f{T_j}{T} \r)^4\\
&&\mbox{Bose} \hspace*{0.75in}\mbox{Fermi}\\
&&\mbox{species} \hspace*{0.6in}\mbox{species}
 \eearrs
As $T$ continues to drop species with rest masses $m_i \gg T$
become nonrelativistic and stop contributing to the above. For
such species,
\bearrs \rho^{non-relat.}_i &=& m_in_i\\[10pt]
&=& \f{m_i\; \l( n_i \l( t_1 \r) R^3 \l( t_1 \r)
\r)}{R^3(t)} \\[10pt]
p^{non-relat}_i&=& 0 \; \rar \; \mbox{``dust"} 
\eearrs

\subsection{Isentropic Expansion}

 We assume particle physics time scales to remain far shorter than
 expansion time scale $H(t)^{-1}$. Thus we have approximate
 equilibrium throughout and we have been using equilibrium
 thermodynamics. Specifically entropy is conserved. While the usual
 term for such a system is adiabatic, the term in the sense of being
isolated from   ``other" systems does not apply to the Universe and we shall
refer directly to fact that entropy is conserved.
With no other system with which to exchange energy, it can be shown for
 the primordial contents of the Universe, that (Kolb \& Turner, Section 3.4)
 \[
 d \l[ \f{( \rho + p)V}{T} \r] = 0 = dS
 \]
 we define $s \equiv \f{S}{V} = \f{\rho +p}{T}$ which is dominated by
 contribution of relativistic particles
 \bearrs
 s&=& \f{2\pi^2}{45} g_{*s} T^3\\[10pt]  
\mbox{with}& &\\
 g_{*s}&=&{\dis \sum_i} g_i \l( \f{T_i}{T}\r)^3 + \f{7}{8} {\dis
 \sum_j} g_j \l( \f{T_j}{T} \r)^3\\
 &&\mbox{Bose}\hspace*{0.72in} \mbox{Fermi}
 \eearrs
Note that this is a different definition of the effective number of 
degrees of freedom than in the case of energy density $\rho$.

\subsection{Temperature Thresholds}
While the conditions of approximate equilibrium and isentropic expansion 
hold for the most part, crucial energy thresholds in microphysics alter the
conditions in the early Universe from time to time and leave behind imprints
of these important epochs. Based on the scales of energy involved, we 
present here a short list of  important epochs in the history of the early Universe

\begin{tabular}{lp{2in}l}
$T$&Relevant species&$g_*$\\[12pt]

$T\ll$ MeV& 3$\nu$'s; photon ( $\gamma$) &3.36 (see reason below)\\[12pt]

1 MeV to 100 MeV& 3$\nu$'s; $\gamma$; $e^+e^-$&10.75\\[12pt]

$\geq$ 300 GeV&8 gluons, 4 electroweak gauge bosons; quarks \& leptons (3
generations), Higgs doublet&106.75\\
\end{tabular}

 Above 300 GeV scale, no known particles are non-relativistic, and
we have the relations
 \bearrs
 H&=& 1.66 g_*^{1/2} \f{T^2}{M_{Pl}}\\[10pt]
 t&=& 0.30l \; g_*^{1/2} \f{M_{Pl}}{T^2} \sim \l( \f{T}{MeV}
 \r)^{-2} sec
 \eearrs
 where
 \[
 G = \f{1}{\l( M_{Pl} \r)^2} = \f{1}{\l( 1.2211 \times 10^{19}
 GeV \r)^2}
 \]

\subsection{Photon Decoupling and Recombination}
\label{sec:decoupling}
We now see in greater detail how one can learn about such thresholds,
with the example of photons. As the primordial medium cools, at some
epoch, neutral hydrogen forms and photons undergo only elastic
(Thomson) scattering from that point onwards. Finally photons decouple from 
matter when their mean free path grows larger than
Hubble distance $H^{-1}$.
 \[
 \Gamma_\gamma = n_e \sigma_T < H
 \]
Here $\sigma_T = 6.65 \times 10^{-25} cm^2$ is the Thomson 
cross-section. (Verify that in eV units, $\sigma_T=1.6\times 10^{-4}\textrm{(MeV)}^{-2}$). 
Note that we should distinguish this event from the process
of ``recombination", which is the formation of neutral Hydorgen. 
In this case the competition is between cross section for ionisation
and the expansion rate of the Universe. 
By contrast the above relation determines the epoch of decoupling of 
photons, also sometimes called ``the surface of last scattering".
Thus the process of recombination ends with residual ionisation which is
the $n_e$ required above to determine decoupling.

\subsubsection{Saha Equation for Ionization Fraction}
To put above condition to use, we need $n_e$ as a function of time or 
temperature and then $H$ at the same epoch. 
$n_e$ should be determined from detailed treatment of non-equilibrium 
processes using Boltzmann equations, which we shall take up later. However, 
utilizing various conservation laws, we can obtain a relationship between
the physical quantities of interest, as was done by Saha first in the 
context of solar corona.

We introduce the number densities $n_H$, $n_p$, $n_e$ of neutral Hydrogen,
protons and neutrons respectively. The amount  of Helium formed is relatively
small, $n_{He} \simeq 0.1 n_p$ and is ignored. Charge neutrality requires 
$n_e = n_p$, and Baryon number conservation requires $n_B = n_H + n_p$.
From approximate thermodynamic equilibrium, we expect these densities to
be determined by the Boltzmann law, where $i$ stands for $H$, $p$ or $e$ :
 \[
 n_i = g_i \l( \f{ m_iT}{2\pi} \r)^{3/2} \exp \l( \f{\mu_i
 -m_i}{T} \r)
 \]
with $g_i$ degeneracy factors, $m_i$ the relevant masses and $\mu_i$
the relevant chemical potentials respectively.  Due to chemical equilibrium, 
\[ \mu_H = \mu_e + \mu_p
\]
Thus we can obtain a relation
 \[
 n_H = \f{g_H}{g_pg_e} n_pn_e \l( \f{m_eT}{2\pi}\r)^{-3/2} \l(
 \f{m_H}{m_p}\r)^{3/2} \exp \l( \f{B}{T} \r)
 \]
where  $B$ denotes the Hydrogen binding energy, $B= m_p+m_e-m_H = 13.6 \; eV$.
Finally we focus on the fraction $X_e \equiv \f{n_p}{n_B}$ of charged baryons 
relative to total baryon number,
 \[
 \f{1 - X^{eq}_e}{\l( X_e^{eq}\r)^2} = \f{4\sqrt{2}}{\sqrt{\pi}}
 \xi (3) \l( \f{n_B}{n_\gamma} \r)\l( \f{T}{m_e}\r)^{3/2} \exp \l(
 \f{B}{T}\r)
 \]
\begin{figure}[htbp]
{\par\centering \resizebox*{0.8\textwidth}{!}
{\rotatebox{0}{\includegraphics{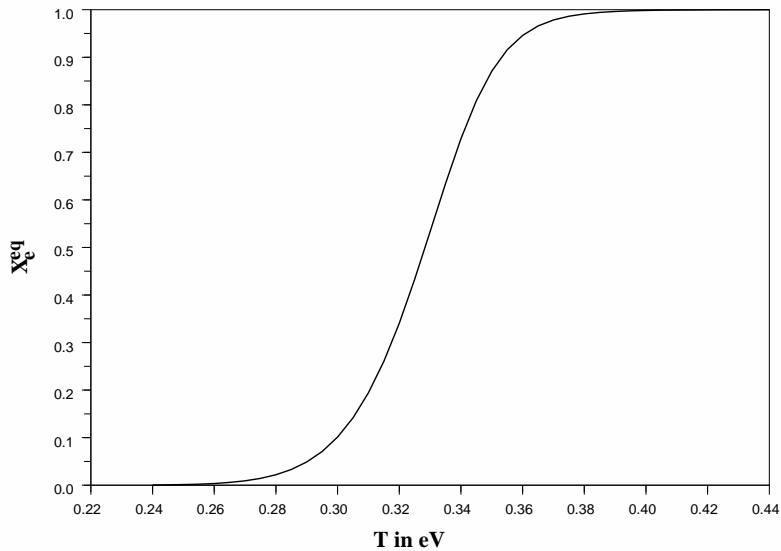}}} \par}\vspace{0.3cm} 
\caption{ Ionization fraction as a function of photon temperature expressed 
in eV}
\label{fig:photon_decoupling}
\end{figure}

Fig. \ref{fig:photon_decoupling} shows variation of $X^{eq}_e$ as function of the
photon temperature $T$, using the value of $n_B/n_\gamma=6\times 10^{-10}$
as known from Big Bang Nucleosynthesis calculations and WMAP data. We see that 
the ionization fraction reduces to
$5$\% when $T\approx 0.29$eV$\approx 350 K$. These are our estimates of the
ionisation fraction (set arbitrarily at 5\%) and the temperature of the last scattering.

Using this estimate we can now calculate the Hubble parameter in the
inequality which determines decoupling. We assume that the Universe was already 
matter dominated by this epoch. Then the scale factor obeys 
\[ R(t)=\frac{R_0}{t_0^{2/3}}t^{2/3}
\]
where $t_0$ and $R_0$ are a particular epoch and the corresponding scale
factor, chosen here to be the current values signified by subscript $0$.
Using this we try to determine the $H$ at decoupling epoch in terms of $H_0$.
\[
\frac{\dot R}{R}=\frac{2}{3 t} = \frac{H}{H_0}H_0 = \l(\frac{\rho}{\rho_0}\r)^{1/2}H_0
=\l(\frac{n_B}{n_{B0}}\r)^{1/2}H_0
\]
Combine this with the relation
\[
\Omega_B=\frac{\rho_{B0}}{\rho_{\textrm{crit}0}}=\frac{m_pn_{B0}}{3H_0^2/8\pi G}
\]
where $m_p$ is the proton mass. Then the condition for photon decoupling becomes 
\[
X_e^{eq}\, \sigma_T < \l(\frac{T_0}{T} \r)^{3/2} \frac{8\pi}{3}G\frac{m_p}{\Omega_BH_0} 
\]
Substituting $\sigma_T$ expressed in eV units computed above, and $X^{eq}_e$ of 5\% 
and the corresponding temperature found from the
graph of Saha formula, we can check that the temperature of the photons
decoupled at that epoch should today be 
\[
T_0 \sim (32 K)\times (\Omega_B)^{2/3} 
\]
Although very crude, this estimate is pretty close to the order of magnitude
of the temperature of the residual photons today. We have thus traced a possible way
Alpher and Gamow could anticipate the presence of residual radiation
from the Big Bang at approximately such a temeperature. 

\section{Phase Transitions with Quantum Fields in the 
Early Universe} 

We have been considering an expanding Universe, but its expansion rate is so 
slow that for most of its history, it is quasi-static as far as Particle
Physics processes are concerned. Under this assumption, we can think of a
thermal equilibrium for
substantial periods on the microscopic scale, and build in the slow
evolution of the scale factor as a significant but separate effect.

Quantized fields which are systems of large numbers of degrees of freedom 
display different collective behavior, with qualitative changes in
ground state as a function of temperature. 
The technique for studying the ground state behavior at non-zero temperature 
is a modification of the process of calculating Green functions using 
path integral method in functional formalism.

\subsection{Legendre Transform}
As a warm up for the method to be used, consider a magnetized spin system
with spin degrees of freedom $s(x)$ as a function of position, and $\cal H$ 
the Hamiltonian of the system.
If we want to determine the equilibrium value of magnetization, we first 
introduce an external field
$H$,  in the presence of which, the Helmholtz Free Energy $F$ is given by
\[
Z(H) \equiv e^{-\beta F(H)} = \int {\cal D}s \exp \l( -\beta \int dx \;
{\cal H}(s) - Hs(x)\r)
\]
where $\beta=1/T$, (same as $1/(kT)$ in our units) and ${\cal D}s$ denotes 
functional integration over $s(x)$.
Now the quantity of interest is 
\[
M = \int dx \langle s(x)\rangle = \f{1}{\beta} \f{\p}{\p H} \log \; Z = -
\f{\p F}{\p H}
\]
We now introduce a function of $M$ itself, the Gibbs free energy, through 
the Legendre transform
\[
G(M) = F + MH
\]
so that by inverse transform,
\[
H = \f{\p G}{\p M}
\]
Now $H$ being an auxiliary field has to be set to zero. Thus the last equation 
can be now read as follows. The equilibrium value of $M$ can be found 
by minimizing the function $G(M)$ with respect to $M$. In other words, for 
studying the collective properties  of the ground state, $G$ is the more 
suitable object than $\cal H $.

\subsection{Effective action and effective potential}
In Quantum Field Theory, a similar formalism can be set up to study
the collective behavior of a bosonic field $\phi$. It is possible in analogy
with the above, to define a functional $\Gamma$ of an argument suggestively
called $\phi_{cl}$ designating the c-number or the classical value of the 
field. Analgous to the external field $H$ above, an auxiliary external current $J(x)$
is introduced. Then
\[
Z[J] \equiv e^{-iE[J]} = \int {\cal D} \phi \exp \l[ i \int d^4x \l(
{\cal L} [\phi ] + j\phi \r) \r]
\]
Then we obtain the relations
\bearrs
\f{\delta E}{\delta J} = i \f{\delta \log Z}{\delta
J(x)} &=& - \langle \Omega
| \phi (x) | \Omega \rangle_J\\[10pt]
&\equiv& \phi_{cl}(x) 
\eearrs
Therefore, let 
\[ \Gamma\l[ \phi_{cl}\r] \equiv - E[J] - \int d^4 y
J(y) \phi_{cl} (y)\] 
so that
\[
\f{\delta}{\delta \phi_{cl}} \Gamma \l[ \phi_{cl} \r] = - J(x)
\]
Thus the quantum system can now be studied by focusing on a classical field
$\phi_{cl}$, whose dynamics is determined by minimizing the functional $\Gamma$.
The auxiliary current $J$ is set zero at this stage. This is exactly as in classical
mechanics, minimizing the action for finding Euler-Lagrange equations. The functional 
$\Gamma$ is therefore called the effective action. If it can be calculated, it  
captures the entire quantum dynamics of the field $\phi$ expressed in terms of the 
classical function $\phi_{cl}$.

Calculating $\Gamma$ can be an impossible project. A standard simplification
is to demand a highly symmetric solution. If we are looking for the properties of
a physical system which is homogeneous and in its ground state, we need the collective 
behavior of $\phi$ in a state which is both space and time translation invariant. 
In this case $\phi_{cl} (x) \rar \phi_{cl}$, $\p_\mu\phi_{cl} =0$ and $\Gamma$ 
becomes an ordinary function (rather than a functional) of $\phi_{cl}$. 
It is now advisable to factor out the spacetime volume to define
\[
V_{eff} \l( \phi_{cl} \r) = - \f{1}{(VT)} \Gamma \l[ \phi_{cl} \r]
\]
so that the ground state is given by one of the solutions of 
\[
\f{\p}{\p\phi_{cl}} V_{eff} \l( \phi_{cl} \r) = 0
\]
In general $V_{eff}$ can have several extrema. The minimum with 
lowest value of $V_{eff}$, if it is unique, characterises the ground state 
while the other minima are possible metastable states of the system. A more interesting
case arises when the lowest energy minimum is not unique. In a quantum system with 
finite number of degrees of freedom, this would not result in any ambiguity. 
The possibility of tunneling between the supposed equivalent vacua determines a 
unique ground state -- as in the example of ammonia molecule.
But in an infinite dimensional system, such as a field system, we shall see that
such tunneling  becomes prohibitive, and the energetically equivalent vacua can 
exist simultaneously as possible ground states.

\begin{figure}[htbp]
{\par\centering \resizebox*{0.7\textwidth}{!}
{\rotatebox{0}{\includegraphics{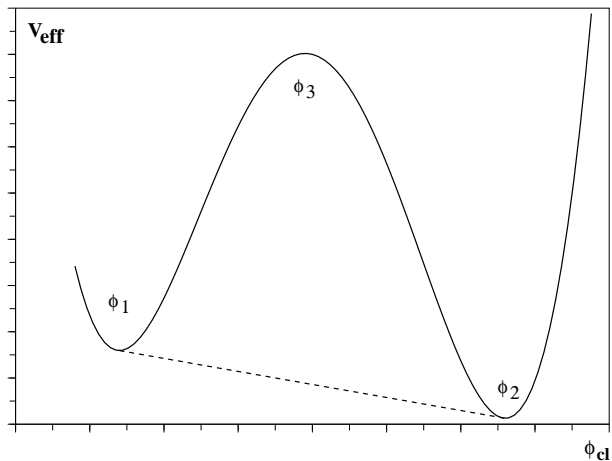}}} \par}\vspace{0.3cm} 
\caption{The effective potential of a system with local minimum at $\phi_1$ and $\phi_2$.
The mostly concave segment with the point $\phi_3$ is treated as unphysical and replaced 
by the dashed line.}
\label{fig:maxwell}
\end{figure}

When we speak of several local minima of $V_{eff}$ and therefore maxima separating
them we are faced with a point of principle. Any function defined as a Legendre
Transform can  be shown to be intrinsically convex, i.e., it can have no maxima, only
minima. The maxima suggested by above extremization process have to be replaced
by a construction invented by Maxwell for thermodynamic equilibria. Consider two
minima $\phi_1$ and $\phi_2$ separated by a maximum $\phi_3$ as shown in \ref{fig:maxwell}. 
We ignore the part
of the graph containing point $\phi_3$ as of no physical significance, and introduce 
a parameter $x$ which permits continuous interpolation between the two minima. 
Over the domain intermediate between  $\phi_1$  and $\phi_2$, we introduce
a parameter $x$ and redefine $\phi_{cl}$ to be 
\[ \phi_{cl}= x\phi_1 + (1-x) \phi_2 \qquad 0 \leq x \leq 1
\]
The assumption is that the actual state of the system is no longer translation invariant
and is an admixture of phases with either a value $\phi_1$ or a value $\phi_2$.
The above redefinition is interpreted as a value of $\phi$ averaged over such regions.
Then if the physical value of $\phi_{cl}$ is characterised by parameter value $x$, 
we define the corresponding  value of $V_{eff}$ 
\[
V^{avg}_{eff} \l( \phi_{cl} \r) = x V_{eff} \l( \phi_1 \r) + (1-x)
V_{eff} \l( \phi_2 \r)
\]
These formulae will not be of direct relevance to us. However we should remember
that while the minima $\phi_1$ and $\phi_2$ represent possible physical situations,
the extremum does not. We shall see that the admixture phase arises when the system
begins to tunnel from one vacuum to another due to energetic considerations. We shall
develop the formalism for the tunneling process and will be more interested in
the tunneling rate which in turn is indicative of how quickly the transition
from a false vacuum (a local minimum) to a true vacuum (absolute minimum)  can be completed.

\subsection{Computing $V_{eff}$}  

A very simple example of a non-trivial effective potential is given by 
a complex scalar field theory with a non-trivial self-coupling
\[
\mathcal{L}= \p_\mu \phi^{\dag} \p^\mu \phi - \frac{\lambda}{4}(\phi^{\dag}\phi-\mu^2 )^2
\]
Note that the mass-squared is negative and so the kinetic terms without
the $\phi^4$ interaction would imply tachyonic excitations. The correct
field theoretic interpretation is to work around a stable minimum such as
$\phi=\mu$. Thus we define $\phi(x)=\mu+\tilde{\phi}(x)$ and quantize only
the $\tilde{\phi}(x)$ degrees of freedom. By substituting the redefined $\phi$
into the above Lagrangian we find that the field $\tilde{\phi}$ has a
mass-squared $+\mu^2$ and can be treated perturbatively.

In this example the polynomial $\frac{\lambda}{4}(\phi^{\dag}\phi-\mu^2 )^2$ 
serves as the zeroth order effective potential. The values $\mu
e^{i\alpha}$ for real values of $alpha$ between $0$ and $2\pi$ are all 
permissible vacuum expectation values (VEVs) as determined from minimizing 
this polynomial. These values are modified in Quantum Field Theory, however
in this simple example their effect will only be to shift the value 
$\mu$ of $|\phi|$ by corrections of order $\hbar$. 
This example is the well known Goldstone mechanism of symmetry breaking.
Here the symmetry breaking is indicated by the classical $V$ itself.

There are important exceptions to this, where quantum corrections
become important and they must be computed. One is where the above
Lagrangian is modified as follows. Remove the mass term altogether, but couple
the massless field to a $U(1)$ gauge field, in other words massless
scalar QED. The classical minimum of the potential $(\phi^{\dag}\phi)^2$
is $\phi=0$. It was shown by S. Coleman and E. Weinberg that after
the one-loop quantum effects are taken into account, the minimum of the
scalar potential shifts away from zero. The importance of this result is
that the corrected vacuum does not respect $U(1)$ gauge invariance.
The symmetry breakdown is not encoded in classical $V$ and has to 
be deduced from the $V_{eff}$.

Another important example is effects of a thermal equilibrium.
Consider the complex scalar of our first example, with mass-squared 
negative at classical level. Suppose we couple this field to a gauge
field. This would lead to spontaneous but classically determinable
breakdown of the gauge invariance according to Higgs-Kibble mechanism.
If we now include temperature corrections, then in the high temperature 
limit the minimum of the scalar field Lagrangian in fact shifts to
$\phi=0$, thus restoring the symmetry! This is the case important
to the early Universe where Universe cools slowly from very high values
of temperature, possibly close to the Planck scale.  Since Higgs
mechanism is the chief device for constructing Grand Unified Theories
(GUTs) we find that gauge symmetries are effectively restored at
high temperatures and non-trivial vacua come into play only at
lower temperatures.

We will be unable to discuss the formalism for computing the effective 
potential in any detail. However here we shall briefly recapitulate 
the recipe which reduces the problem to that of computing Feynman 
diagrams. Recall that an important consequence of quantum corrections 
can be change in the vacuum expectation value of a scalar field.
We shall assume that the vacuum continues to be translation invariant,
so that this shifted value is a constant. 
Anticipating this, in the Lagrangian we shift the field value
$\phi(x) = \phi_{cl} + \eta (x)$
where the significance of the subscript $cl$ becomes clear due
to the reasoning given above. However note that at his stage 
$\phi_{cl}$ is only a parameter and the $V_{eff}$ is computed as
a function of it. Now in the presence of an external source $J(x)$,
\[
\barr{ll} {\dis \int} d^4x \l( {\cal L} + J \phi \r) &= {\dis
\int} d^4 x \l( {\cal L} + J \phi_{d} \r) + {\dis \int} d^4x \eta
\l(
\f{\delta{\cal L}}{\delta\phi} + J \r)\\[10pt]
&+ \f{1}{2} {\dis \int} d^4x d^4y \eta \; \eta \f{\delta^2 {\cal
L}}{\delta \phi \delta \phi} .... \earr
\]
Here onwards we assume that the $\phi_{cl}$ is chosen to satisfy
\[
\l. \f{\delta{\cal L}}{\delta\phi} \r|_{\phi=\phi_{cl}}= -J(x) 
\] 
Note that this $\phi_{cl}$ is still dependent on the external function
$J(x)$ and hence adjustable. Then in the path integral formula for the 
generating functional, we can carry out a saddle point evaluation
of the gaussian\footnote{Actually ``pseudo''-gaussian due to the presence of $i$, 
but equivalently after choosing to work with euclidian path integral.} 
integral around the extremum defined by the choice of $\phi_{cl}$ just made.
\[
\barr{ll} {\cal Z}_J  &={\dis \int} {\cal D} \eta \exp \l( i {\dis \int} {\cal
L} \l( \phi_{cl} \r) + J\phi_{cl} + \f{1}{2} {\dis \int} \eta {\cal
L}^" \eta  \right) \\[10pt]
& = \exp \l[ i  {\dis \int} {\cal L} \l( \phi_{cl} \r) + J\phi_{cl}
\r] \times \l(  det \l[ - \f{\delta^2 {\cal L}}{\delta \phi \delta
\phi} \r] \r)^{-1/2} {\cal Z}_2\earr 
\]
where the ${\cal Z}_2$ denotes all the terms of order higher in the 
variations of ${\cal L}$ with respect to $\phi$. In perturbative interpretation
hese higher derivatives
\[
 \f{\delta^n{\cal L}}{\delta \phi \ldots \delta\phi_n}
\]
are treated as vertices, while
\[
-i \l( \f{\delta^2{\cal L}}{\delta\phi\delta\phi}\r)^{-1}
\]
is used as propagator. The main correction to one-loop order however can
be determined directly from the determinant resulting from saddle point
integration. While calculating $S$-matrix elements this determinant is only
an overall constant and of no significance. But in the effective potential
formalism it provides the main correction. It also requires proof to know that
the determinant provides the leading correction and that everything inside
${\cal Z}_2$ represents higher order quantum corrections. This will not be
pursued here. We now quote an example.

\subsubsection{An example}
Consider the theory of two real scalar fields $\phi_1$ and $\phi_2$
\[
{\cal L} = \f{1}{2} \sum_i \l( \p_\mu \phi^i \r)^2 + 
\f{1}{2} \mu^2 \sum_i \l(\phi^i \r)^2 - \f{\la}{4} \l[\sum_i \l( \phi^i \r)^2 \r]^2
\]
with $i=1,2$ and $\mu^2>0$. The latter condition means that the
classical minimum of the theory is not at $\phi_i=0$ but at any of
the values defined by $\lambda(\phi_1^2+\phi_2^2)=\mu^2$. A possible
minimum is at $(\phi_1,\phi_2)$ $=(\mu/\sqrt{\lambda},0)$. If we shift 
the fields by choosing $(\phi_1,\phi_2)$ $=(\phi_{cl}+\eta_1(x),\eta_2(x))$ 
we get propagators for the two real scalar fields $\eta_i$ with
mass-squares given by $m_1^2=3\lambda\phi_{cl}^2-\mu^2$ and 
$m_2^2=\lambda\phi_{cl}^2-\mu^2$.

Evaluation of the determinants of the inverse propagators requires
a series of mathematical tricks.
\bearrs
\log \det \l( \p^2 + m^2\r)&=& Tr \log \l( \p^2 + m^2
\r)\\[10pt]
&=& {\dis \sum_k} \log \l( -k^2 + m^2 \r)\\[10pt]
&=& (VT) {\dis \int} \f{d^4k}{(2\pi )^4} \log \l( -k^2+m^2\r)
\eearrs
The determinant actually has a diagrammatic interpretation as was
explained by Coleman and Weinberg. Consider a single closed loop 
formed by joining together $n$ massless propagators, joined together
by consecutive mass insertions. Here we are treating treating mass 
as an interaction for convenience. This loop has $\frac{1}{n!}$ in
front of it from perturbation theory rules due to $n$ mass insertions.
But there are $(n-1)!$ ways of making these identical mass isertions.
This makes the contribution of this loop weighed by $\frac{1}{n}$.
Further because the propagators are identical, $n$th term is
$n$th power of first term. Thus summing all the terms with single
loop but all possible mass insertions amounts to a $\log$ series.
The next important argument is that indeed the perturbative expansion 
is an expansion ordered by the number of loops. By including all
contributions at one loop, we have captured the leading quantum
correction.

We now turn to evaluation of this using dimensional regularization prescription
\[
{\dis \int} \f{d^dk}{\l( 2\pi \r)^d} \log \l( -k^2 + m^2 \r) =
\f{-i \Gamma \l( - \f{d}{2} \r)}{(4\pi )^{d/2}} \f{1}{\l(
m^2\r)^{-d/2}}
\]
Therefore,
\bearrs
V_{eff}\l( \phi_{cl} \r)&=& - \f{1}{2} \mu^2 \phi^2_{cl} +
\f{\la}{4} \phi^4_{cl}\\[10pt]
&&- \f{1}{2} \f{\Gamma \l( - d/2\r)}{(4\pi )^{d/2}} \l[ \l(
\la \phi^2_{cl} - \mu^2\r)^{d/2} + \l( 3\la \phi^2_{cl}-\mu^2\r)^{d/2}
\r]\\[10pt]
&&+\f{1}{2} \delta_\mu \phi^2_{cl} + \f{1}{4} \delta_\la \phi^4_{cl}
\eearrs
where the $\delta_\mu$ and $\delta_\la$ represent finite parts.
Now
\bearrs
\f{\Gamma\l( 2 - d/2\r)}{(4\pi )^{d/2} \l( m^2
\r)^{2-d/2}} &=& \f{1}{(4\pi )^2} \l( \f{2}{\epsilon} - y + \log
4\pi -
\log m^2 \r)\\[10pt]
&&\xrightarrow{\overline{\mbox{MS}} \mbox{ scheme}} \f{1}{(4\pi
)^2} \l( -\log \f{m^2}{M^2} \r) 
\eearrs
where a reference mass scale $M$ has to be introduced.
Therefore, with superscript $(1)$ signifying one-loop correction,
\bearrs
V^{(1)}_{eff}&=& \f{1}{4} \f{1}{(4\pi )^2} \l[
\l( \la \phi_{cl}^2 - \mu^2 \r)^2 \l( \log \l( \la \phi^2_{cl} - \mu^2
\r)/M^2-3/2\r)\r]\\[10pt]
&&+ \l( 3\la \phi^2_{cl}-\mu^2\r)^2 \l( \log \l[ \l( 3\la \phi^2_{cl} -
\mu^2 \r)/M^2\r] - 3/2 \r) 
\eearrs
A rule of thumb summary of this example is that the one-loop
correction to the effective potential is $(1/64\pi^2)$
$m_{eff}^4\ln (m_{eff}/M)^2$
where $m_{eff}^2$ contains $\phi_{cl}^2$ and $M$ is a reference
mass scale required by renormalization. Further, it can be shown 
that the modification to the tree level (classical) vacuum
expectation value comes only from the $m^4\ln (m/M)^2$ term of
the field direction in which the vacuum is already shifted, $\phi_1$ 
in the present example. 

\subsection{Temperature  corrections to 1-loop}
In the early universe setting we are faced with doing Field Theory
in a thermal bath, also referred to as ``finite temperature field theory''. 
With some clever tricks and exploiting the analogy of the field theory
generating functional with the partition function for a thermal ensemble,
one can reduce this problem also to that of calculating an effective
potential. The key modification introduced is to combine the time
integration of the generating functional and the multiplicative
$-\beta$ (inverse temperature) occurring in the partition function 
into an imaginary time integral $-\int_0^\beta d\tau$. Further,
the trace involved in the  thermal averaging can be shown to be 
equivalent to periodicity in imaginary time of period $\beta$. 
For bosonic fields one is lead to periodic boundary condition and 
for  fermionic fields one is lead to anti-periodic boundary condition.
Thus the usual propagator is replaced by an imaginary time
propagator of appropriate periodicity. For a scalar field of mass 
$m$, thermal propagator $\Delta^T$ is given by
\[
\Delta^T(x,y)\ =\ \frac{1}{\beta}\sum_{k^0=2\pi in/\beta} 
\int \frac{d^3k}{(2\pi)^3}e^{-i k\cdot(x-y)}\frac{i}{k^2-m^2}
\]

\begin{figure}[htbp]
{\par\centering \resizebox*{0.7\textwidth}{!}
{\rotatebox{0}{\includegraphics{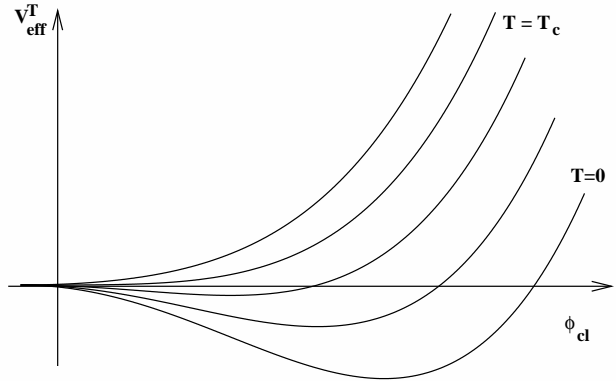}}} \par}\vspace{0.3cm} 
\caption{ Temperature dependent effective potential plotted to show its
dependence on $\phi_{cl}$ for various values of temperature $T$. 
$T_c$ denotes the temperature below which  the trivial minimum is unstable. } 
\label{fig:VeffT}
\end{figure}

We now quote the result for the temperature dependent effective
potential $V^T$. We focus only on the $\phi_1$ degree of freedom
of the previous example and drop the subscript $1$,
\bearrs
V^T_{eff} \l[ \phi_{cl} \r]&=& V_{eff} \l[ \phi_{cl} \r] +
\f{T^4}{2\pi^2} {\dis \int}^\infty_0 dx x^2 \ln \l[ 1 - \exp \l(
- \l( x^2 + \f{m^2}{T^2}\r)^{1/2}\r)\r]\\[10pt]
\mbox{\rm with}\quad m^2\l( \phi_{cl} \r)&=& - \mu^2 + 3\la \phi^2_{cl}
\eearrs
and  $V_{eff}$ to one-loop order is as obtained in the previous
subsection.
In the high temperature limit $T \gg \phi_{cl}$ we can 
determine the leading effects of temperature by expanding the above
expression to find
\[
V^T_{eff} = V_{eff} + \f{\la}{8} T^2 \phi^2_{cl} - \f{\pi^2}{90} T^4 + \ldots
\]
From thermodynamic point of view this $V_{eff}$ represents the 
Gibbs free energy. Entropy density, pressure and the usual energy 
density are given by
\bearrs
s[\phi_{cl}] &=& - \f{\p V^T_{eff}}{\p_T}; 
\qquad \textrm{while}\; p = -V^T[\phi_{cl}]\\[5pt] 
\textrm{and}  \rho[\phi_{cl}] &=& V^T_{eff} + T s[\phi_{cl}]\\ [5pt]
  &=& V \l( \phi_{cl} \r) -
\f{\la}{8}T^2 \phi^2_{cl} + \f{\pi^2}{30} T^4.
\eearrs
The resulting graphs of $V^T_{eff}$ are plotted in fig. \ref{fig:VeffT}.
In plotting these, the term $T^4$ which is the usual thermodynamic contribution, 
but which is independent of the flield $\phi_{cl}$ is subtracted.

We can summarise the main results of this subsection retaining
a single scalar degree of freedom $\phi$ for which,
\[
 {\cal L}= \f{1}{2} \p_\mu \phi \p^\mu \phi + \f{1}{2} \mu^2
 \phi^2 - \f{1}{4} \la \phi^4 
\] 
with minima at $\sigma_\pm = \pm \sqrt{\mu^2/\la}$

The leading effect of the temperature correction is to add 
a term $T^2\phi_{cl}^2$ so that
 \[
 V^T_{eff} = \l( - \f{1}{2} \mu^2 + \f{\la}{8} T^2 \r) \phi^2_c +
 \f{1}{4} \la \phi^4_c + \ldots
 \]
 This is extremized at the values
 \[
\phi_{cl} = \pm \sqrt{\f{\mu^2 - \la T^2/4}{\la}} \mbox{ and } \phi_{cl} = 0
\]
Further, note that
\[
 \f{\p^2V}{\p\phi^2}\mid_{\phi_c=0} = -\mu^2 + \f{\la}{4} T^2
 \]
As a result we see that the curvature can change sign at the
trivial extremum $\phi_{cl}=0$ at a critical temperature $T_c=
2\mu/\sqrt{\la}$.

\section{First order Phase transition}
At the end of the last section we saw that the minimum at $\phi=0$
can turn into a maximum as the temperature drops below the critical
value $T_c$. This effect is felt simultaneously throughout the system
and a smooth transition to the newly available minimum with $\phi\neq 0$
ensues. The expectation value of $\phi$ is called the \textit{order 
parameter} of the phase transition, and in this case where it changes 
smoothly from one value to another over the entire medium is called 
a Second Order phase transition.

But there can also be cases where there are several minima of the free 
energy function but separated by an energy barrier. Thus the system 
can end up being in a phase which is a local minimum, called the
\textit{false vacuum}, with another phase of lower free energy, called 
the \textit{true vacuum}, available but not yet accessed. 
If the barrier is not too high, thermal fluctuations can cause the system 
to relax to the phase of lower free energy. The probability for the system
to make the transition is expressed per unit volume per unit time
and has the typical form 
\[
\Gamma = A \exp\{-B \}
\]
The expressions $A$ and $B$ are dependent upon the system under
consideration. The presence of $B$ reminds us of the Boltzmann
type suppression that should occur if the system has to overcome
an energy barrier in the process of making the transition. 
In Quantum Field Theory there are also quantum mechanical 
fluctuations which assist this process. We observe that the formula
above is also of the type of WKB transition rate in Quantum Mechanics.
Indeed, we have partly thermal fluctuations and partly tunneling 
effects responsible for this kind of transition. A convenient
formalism exists for estimating the combined effects using the
thermal effective potential introduced in the previous section.

A transition of this type does not occur simultaneously over the
entire medium. It is characterized by spontaneous occurrance of
small regions which tunnel or fluctuate to the true vacuum. Such 
regions of spontaneously nucleated true vacuum are called ``bubbles'' 
and are  enclosed from the false vacuum by a thin boundary called 
the ``wall''. Since the enclosed phase is 
energetically favorable, such bubbles begin to expand, 
as soon as they are formed, into the false vacuum.
Over time such bubbles keep expanding, with additional bubbles continuing 
to nucleate, and as the bubbles meet, they merge, eventually completing 
the transition of the entire medium to the true vacuum. Such a 
transition where the order parameter $phi$ has to change abruptly 
from one value to another for the transition to proceed is called 
a First Order phase transition (FOPT). 

\subsection{Tunneling}
\begin{figure}[htbp]
{\par\centering \resizebox*{0.7\textwidth}{!}
{\rotatebox{0}{\includegraphics{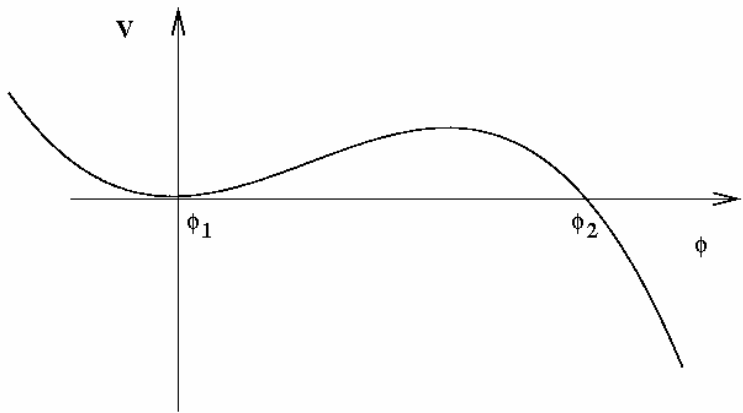}}} \par}\vspace{0.3cm} 
\caption{A system which has a local minimum at $\phi_1$ and which is 
unstable towards tunneling to a point $\phi_2$ of equal energy }
\label{fig:Vbarrier}
\end{figure}

At first we shall consider tunneling for a field system only
at $T=0$. An elegant formalism has been developed which 
gives the probability per unit volume per unit time for the formation 
of bubbles of true vacuum of a given size. Consider a system
depicted in fig. \ref{fig:Vbarrier} which has a local minimum
at value $\phi_1$, chosen to be the origin for convenience.
There are other configurations of same energy, such as $\phi_2$
separated by a barrier and not themselves local minima.
If this is an ordinary quantum mechanical system of one variable
and the initial value of $\phi$ is $\phi_1$, the system is 
unstable towards tunneling to the point $\phi_2$ 
and subsequently evolving according to usual dynamics. 
If this were point particle mechanics, the formula for the transition 
amplitude from the state $\vert\phi_{1}\rangle$ to
the state $|\phi_{2}\rangle$ is given in the Heisenberg picture and
in the path integral formulation as
\[
\langle\phi_{2}\vert e^{-\frac{i}{\hbar}HT}\vert\phi_{1}\rangle
=\int\mathcal{D}\phi e^{\frac{i}{\hbar}S}
\]
where $T$ is a time interval and the action
$S$ on the right hand side has the same range of time integration.
Instead of evaluating this directly, we make two observations. Firstly,
if we asked for the amplitude for the state $\vert\phi_{1}\rangle$
to evolve into itself after time $T$, it would involve contributions
also from paths that access the state $\vert\phi_{2}\rangle$. Thus
if we inserted a complete set of states on the left hand side above
at an intermediate time, say, time $T/2$ (we justify the $T/2$
later), among the many contributions there would also occur the term 
\[
\langle\phi_{1}\vert e^{-\frac{i}{\hbar}HT/2}\vert\phi_{2}\rangle
\langle\phi_{2}\vert e^{-\frac{i}{\hbar}HT/2}\vert\phi_{1}\rangle
\]
Correspondingly on the right hand side we would have contribution
from paths that start at the point $\phi_{1}$ and end at $\phi_{1}$,
but after reaching $\phi_{2}$ somewhere along the trajectory.

The second point is more interesting. Actually the presence of 
$\vert\phi_{2}\rangle$, a state of equal energy, makes 
$\vert\phi_{1}\rangle$ unstable. Hence the contribution such as 
$\langle\phi_{1}\vert e^{-\frac{i}{\hbar}HT/2}\vert\phi_{2}\rangle$ 
actually makes the total amplitude for $\phi_{1}$ returning to $\phi_{1}$
smaller than unity in magnitude. This happens only if the evolution
operator $e^{-\frac{i}{\hbar}HT/2}$ somehow departs from being of
unit magnitude, i.e., its exponent becomes real negative rather than
pure imaginary.

Thus if we look, not for the entire amplitude, but only for the part
where the exponent becomes effectively imaginary then that part of
the sum over intermediate states actually indirectly gives the 
transition amplitude $\langle\phi_{2}\vert e^{-\frac{i}{\hbar}HT/2}
\vert\phi_{1}\rangle$,
the one we started out to look for (aside from factor $1/2$ in time). 
In the limit $T$ becomes infinite, all the contributions with real 
negative exponents will go to zero. The leading contribution is 
the term with the smallest exponent. On the right hand side this means
that among all the paths that start from one vacuum, sample the other
and return, the one that minimises the action will contribute. Again
we expect, on the right hand side, the exponent to be real negative,
i.e., the contribution of a Euclidean path, with $i\int dt$ replaced
by $-\int d\tau$. This is also reasonable since we know the usual
kinetic energy of the particle has to be replaced by a negative 
contribution when the trajectory is under the barrier.

The summary of this discussion is that actually we should be looking
only for the imaginary part of the contributions on both the sides
of the formula above. If we find the path which minimises
the \textit{Euclidean action}, then in terms of that, to leading order, 
and in the semi-classical limit, we have the tunneling formula
\[
\Gamma = A \exp \l( - S_E \r)
\]
We can also
now see the reason for $T/2$ to be the appropriate time. If the
path minimises the action it should be as symmetric as possible.
Thus we expect a time symmetry $T\rightarrow -T$ and this explains
why the escape point $\phi_2$ should occur at $T/2$. We therefore
solve the Euler-Lagrange equations derived from the action
\[
S_{E}=\int d^{4}x\{\frac{1}{2}\left(\frac{d\phi}{d\tau}\right)^{2}
+\frac{1}{2}\left\vert\nabla\phi\right\vert^{2}+V[\phi]\}
\]
Now the action will be minimum for the path that has the fewest 
wiggles, i.e., is mostly monotonic. We expect the path to start 
at large negative $\tau$ at the value $\phi_{1}$ and stay at 
that value as much as possible, and monotonically reach 
$\phi_{2}$ near the origin, and then retrace a symmetric path 
back to $\phi_{1}$ as $\tau$ goes to infinity. Such a path which 
bounces back has been called {}``the bounce''.

To solve for the bounce, our first simplification will be to 
invoke space-time symmetries, viz., we assume that the 
configuration of fields which will minimise the integral in 
question will obey spatial isotropy. This is same as assuming 
that the spontaneously formed bubble will be spherically symmetric. 
With $4$ Euclidean dimensions, assuming $0(4)$ symmetry, one solves 
the equation 
\[ \phi^" + \f{3}{r}
\phi^\pr - V^\pr (\phi ) = 0
\]
One boundary condition is $\phi (r \rar \infty ) = \phi_1$ where
we have chosen $\phi_1=0$. At the origin we could place the requirement
$\phi(r=0)=\phi_2$ but it is more important to require
$\phi^\pr (r=0)=0$ as is usual for spherical coordinates when the 
solution is expected to be smooth through the origin. Indeed we may 
not even know what the ``exit point'' $\phi_2$ after the tunelling will
be. The bounce when solved for will also reveal it. 
\begin{figure}[htbp]
{\par\centering \resizebox*{0.7\textwidth}{!}
{\rotatebox{0}{\includegraphics{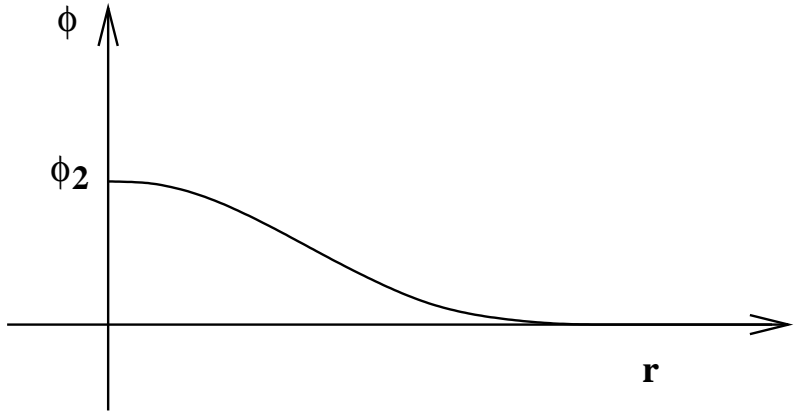}}} \par}\vspace{0.3cm} 
\caption{Tunneling rate is determined by the minimum of the  Euclidean 
action $\phi_{bounce}(r)$ obeying appropriate symmetry and boundary conditions.}
\label{fig:bounce}
\end{figure}
A typical bounce solution is shown in figure \ref{fig:bounce}.

\subsubsection{Next to leading order}
According to above discussion, the tunneling rate is given by a
WKB type formula $\Gamma = A \exp \l( - S_E \r)$. With $S_E(\phi_{bounce})$
determined by extremising the Euclidean action, the exponential is
the most important factor in this formula. The front factor
$A$ arises from integration over the small fluctuations around the 
stationary point $\phi_{bounce}$. This is a Gaussian integration
and results in a determinant. There are several subtleties which arise.
The answer is that we need to remove zero mode(s) of the fluctuation
operator, and normalize with respect to the determinant
of the fluctuations in the absence of the bounce. Thus with prime
on ``det'' denoting removal of zero mode(s),  
\[
A= \l( \f{S_E(\phi )}{2\pi}\r)^2 \l( \f{\det^\prime 
\l[ -\square_E + V^" (\phi_{bounce} )\r]}
{\det \l[ - \square_E + V^"(0
)\r]}\r)^{-1/2} 
\]

\subsubsection{Thermal bounce}
We can now address the tunneling problem in a thermal ensemble,
i.e.,  at $T\neq 0$. Recall our observation during the discussion of 
thermal effective potential, viz., the analogy between Euclidean 
path integral and the trace weighed by density matrix in thermal
partition function. By the same arguments we can show that we must
look for a bounce solution periodic in imaginary time with period
$1/\beta$. There will be an infinite number of such bounces, but the
one to dominate will have least number of extrema. Then we obtain 
the rate formula
\[
\Gamma^{[T]} = A \exp \l( - S_E^{[T]} \r)
\]
with
\[
S_E^{[T]} = 4\pi \beta {\dis \int}^\infty_0 r^2 dr \l[ \f{1}{2}
\phi^{\pr 2} + V^T_{eff}(\phi )\r]
\]
To obtain the euclidian action in this case, we solve the equations
of motion obtained by varying this action and solve them subject 
to $O(3)$ symmetry of spatial directions. This is called the ``bounce''
solution relevant to thermal transitions and the corresponding value
of the action is inserted into the rate formula.

\subsection{Applications}
The formalism developed in this section is important for determining
the evolution of the Universe when the field theory signals a
first order phase transition. The rate $\Gamma$ can vary greatly
due to the exponential factor. If the rate is too small, the expansion 
rate of the Universe may be faster and in this case parts of the
Universe may never tunnel to the true vacuum. If the state of 
broken symmetry  is phenomenologically  desirable, the rate 
should be fast enough, at least faster than the expansion rate
of the Universe at the time of Big Bang Nucleosynthesis (BBN).
The end of a first order phase transition dumps a certain
amount of entropy into the Universe, similar to latent heat in
usual substances.  Since the history of the Universe after the 
BBN is fairly precisely known any unusual phenomenon especially 
one that may disturb the baryon to entropy ratio should occur 
well before the BBN and not alter the required value of the ratio.

Such considerations place constraints on the parameters of 
the scalar field theory undergoing the phase transition. The 
first proposal of inflationary Universe required a fairly 
specific range for the value of the rate, slow enough for 
sufficient inflation to occur, but still fast enough that 
the present day Universe would be in the true vacuum.

In some theories the false vacuum is phenomenologically the 
desirable one. In many supersymmetric models, the desirable state 
actually turns out to be metastable, whereas the true ground 
state has undesirable properties such as spontaneous breaking 
of the QCD colour symmetry. Several models of supersymmetry breaking
arising in a hidden sector and communicated to the observed sector
by messengers, end up with unphysical ground states. In such cases
one invokes the possibility that the parameters of the theory
make the tunneling rate much smaller than the expansion rate 
of the universe till the present epoch. If a volume of the size
$M^{-3}$ determined by the energy scale $M$ of the high energy theory
is to not undergo a transition within the typical expansion
time scale of the Universe, then
\[
\Gamma < M^3H_0
\]
where $H_0$ is the present value of the Hubble parameter,
i.e., the experimentally observed Hubble constant.

As another application, in Standard Model, the Higgs boson has a
self interaction potential of the type discussed above. The exact
form of the potential, namely, which local minimum is energetically 
favored and what can be the tunneling rate for going from one
vacuum to another is determined by the mass parameter and the
quartic coupling occurring in the Higgs potential.
The mass parameter is known from the requirement of spontaneous
symmetry breaking to reproduce the Weak interaction scale. However,
the quartic coupling will be determined only when the collider
experiments will determine its mass. Since this value is as yet
unknown, we can use cosmology to put a bound on its possible values.

It can be shown that if the Higgs boson is very light, then there
is a danger for the Universe to be trapped in an unphysical vacuum.
This puts a lower bound on the Higgs boson mass at about $10$ GeV. 
This is known as the  Linde-Weinberg bound.  

\section{Inflationary Universe}

It is remarkable that the Friedmann-Robertson-Walker model is
so successful a description of the observed Universe. At first
this seems a resounding triumph of General Relativity. It is
true that the dynamics all the way back to Big Bang Nucleosynthesis
(BBN) is successfully described. However one
begins to notice certain peculiarities of the initial conditions.
First of all, the Big Bang itself presents a problem to classical 
physics, being a singularity of spacetime. But we expect this 
to be solved by a successful theory 
of Quantum  Gravity. But now we will show that even the conditions 
existing after  the Big Bang and well within the  realm of 
classical General Relativity pose puzzles and demand a search for 
new dynamics or newer laws of physics.

\subsection{Fine tuned initial conditions}
Any physical entity such as a planet or a star or a galaxy has 
an associated legth or mass scale. Very often this is set by 
accidental initial conditions. However
not all accidentally possible scales would be tolerated by the
dynamics holding together the object. Some initial conditions 
will lead to unstable configurations. Question : what scales 
would be  permitted by a universe driven according to 
General Relativity? 
The coupling constant of Gravity is dimensionful, 
$G_N^{-1}=$ $M^{2}_{Pl}\sim (10^{19}GeV)^{2} $. 
The available physical quantity is energy density.
We propose a naive possibility that the system size or 
scale be decided by this energy density re-expressed in 
the units of $M_{Pl}$. We then find that the Universe has far 
too small an energy density $10^{-6}(eV)^4$  
$\sim 10^{-66}(\mathrm{M_{Pl}})^4$, and far too big 
a size, Giga parsec compared to $10^{-20}$fermi. 
It is interesting that Gravity permits a viable solution
with such variance from its intrinsic scales.

Further, we find that above mentioned ratio evolves 
with time because the scale factor grows and density
keeps reducing. We can ask what is the time scale
set by gravity for this variation. In the evolution equation for the 
scale factor $R(t)$ we see that the intrinsic scale is set
by the Gravitational constant. If the $\rho/(M_{Pl})^4$ was
order unity at some epoch, then there is only one independent time
scale left, that set by $M_{Pl}$.
We then find that the universe would either have recollapsed or 
expanded precisely within the time set by the Planck scale, $10^{-44}$sec. 
The fact 
that the Universe seems to be hovering between collapse and rapid 
expansion even after $14$ billion years, requires that we must start with
extremely fine tuned initial conditions. The fine tuning has to be
to the extent of one part in $10^{66}$ because we started with the 
value of the ratio close order unity. 

It is such uncanny fine tuning that is the motivation for proposing 
an ``inflationary''  event in the early Universe, a phase of unusually 
rapid expansion. Such an event reconditions the ratios we discussed above.
Their large apparant values then arise from dynamics rather than initial 
values.  All physical realizations of this proposal 
have consisted of admitting dynamics other than Gravity to intervene 
for the purpose of significantly  reconditioning these ratios.

It is fully likely that the answer to the puzzles to be described
in the following is buried in the Planck era itself. Any data
which can throw light on such a mechanism would also provide a valuable
window into Planck scale physics. But there are viable candidates 
within the known physical principles of Relativistic Field Theory, 
a possibility which if true would reduce our intrinsic ignorance of 
the physical world, and in turn lead to prediction of newer forces.

\subsection{Horizon problem}
This problem arises from the fact that our Universe had a finite 
past rather than an indefinitely long past. A finite past gives 
rise, at any given time,  to a definite physical size over which 
information could have travelled upto that time. This physical scale 
is called the ``particle horizon''. At present epoch we find unusually 
precise correlation in the physical conditions across many particle
horizons, i.e., over regions of space that had  no reason to be 
in causal contact with each other.
It is to be noted that the world could
well have emerged all highly correlated from the Planck
era. But as explained above, we work in the spirit of 
exploring newer dynamics within the known principles.

With this preamble, the paradox presented by the current
observations is as follows. We know that the observed
Cosmic Microwave Background Radiation (CMBR) originated
at the time of decoupling of photons from the partially 
ionized Hydrogen. The temperature of this decoupling, as
we estimated in sec. \ref{sec:decoupling} is $1200$K,
while today it is close to $1$K. From the ratio of 
temperatures, and assuming a matter dominated Universe,
\[
\f{T_0}{T_{dec}} \simeq \f{1}{1200} = \f{R\l( t_{dec}\r)}{R \l(
t_0\r)} = \l( \f{t_d}{t_0}\r)^{2/3}
\]
Therefore, $t_d \sim 2 \times 10^5 \; h^{-1}$ years.\\

Now consider the size of the particle horizon at these
two epochs, ie., the size of the region over which communication
using light signals could have occurred since the Big Bang.
\[
a \l( t_0 \r) {\dis \int}^{t_0}_{t_d} \f{dt^\pr}{R\l( t^\pr \r)}
\approx 3 t_0 \approx 6000 \; h^{-1} \; Mpc
\]
where the contribution of the lower limit is ignorable.
Similarly at $t_{dec}$, horizon $\sim 3t_{dec} \approx 0.168 \; h^{-1}
Mpc$, using the time-temperature relation appropriate to
the radiation dominated era. Then the angle subtended
to us today by a causally connected region of the decoupling
epoch is
\[
\theta_d = \f{168}{6000} \simeq 0.03 \; rad \;
\mbox{ or 2 deg}
\] 
This means that we are viewing today $\f{4\pi}{(0.03)^2} \approx$ 
$14,000$ causally \textit{un}connected horizon patches, and yet
they show remarkable homogeneity.

\subsection{Oldness-Flatness problem}
An independent puzzle arises due to the fact that the curvature
of the three dimensional space is allowed to be non-zero in 
General Relativity. Let us rewrite the evolution 
equation by dividing out by $H(t)^2\equiv ({\dot R}(t)/R(t))^2$  
\[
\frac{1}{H(t)^2}\frac{k}{R(t)^2} = \Omega(t) - 1 
\]
where $\Omega(t)\equiv 8\pi G\rho(t)/3H(t)^2 $ which can be
thought of as the energy density at time $t$ re-expressed in the units
of $G$ and $H(t)$. At the present epoch $t_0$ in the Universe, the
observational evidence suggests the right hand side (RHS) of above 
equation  is $\pm 0.02$. This suggests the simple possibility that 
the value  of $k$ is actually zero. Let us first assume that it 
is non-zero and assume a power law expansion $R(t)=R_0 t^n$, 
with $n<1$ as is true for radiation dominated and matter dominated 
cases. After dividing the previous equation on both sides
by the corresponding quantities at present epoch $t_0$, we find
\[
\l( \frac{t}{t_0} \r)^{1-n} = 50\times(\Omega(t) - 1)
\]
where we have used the current value of the right hand side (RHS), 
$0.02$. Now the current value $t_0$ is $\approx 5\times 10^{17}$ second,
while at the time of Big Bang Nucleosynthesis (BBN) it was only
about $100$ seconds old. Using $n=1/2$ for the sake of argument, we find
LHS$\approx 10^{-7}$, which means that correspondingly, on the RHS
$\Omega$ must be tuned to unity to one part in $10^{8}$. 
If further, we compare to earlier epochs such as the QCD phase 
transition or the electroweak epoch, we
need higher and higher fine tuning to achieve the $0.02$ accuracy
at present epoch. We thus see that the initial conditions have to
be fine tuned so that we arrive at the Universe we see today.
Equivalently, since the problem is connected to the large ratio
of time scales on the LHS, we may wonder why the Universe has lived so
long. This may be called the ``oldness'' problem.

In case the $k$ is zero, then that would be a miraculous fine 
tuning in itself. The discrete values $0, \pm1$ arise only after
scaling the curvature by a fiducial length-squared. The natural 
values for the curvature pass smoothly from negative to positive 
and zero is only a special point. The tuning of the value to zero
earns this puzzle the name ``flatness'' problem.

We can also restate the problem as there being too much entropy in 
the present Universe. Taking the entropy density 
of the CMB radiation at 2.7K and multiplying by the size of the
horizon as set by approximately $H_0^{-1}\sim (1/3)t_0$ $\sim 3\times 10^9$ yr.
we get the entropy to be $10^{86}$ ( check this!). The aim of inflationary
cosmology is to explain this enormous entropy production as a
result of an unusual phase transition in the early Universe.

\subsection{Density perturbations}
A last important question to be answered by Cosmology is the origin
of the galaxies, in turn of life itself and ourselves. If the Universe
was perfectly homogeneous and isotropic, no galaxies could form.
Current observations of the distribution of several million galaxies
and quasars suggests that the distribution of these inhomogeneities 
again shows a pattern. To understand the pattern one studies the 
perturbation in the density, $\delta\rho(\mathbf{x},D) = \rho(\mathbf{x},D)-{\bar\rho}$, 
where $\bar\rho$ is the average value and $\rho(\mathbf{x},D)$ is the 
density determined in the neighborhood of point $\mathbf{x}$ by averaging
over a region of size $D$. It is found that the density fluctuations 
do not depend on the scale of averaging $D$. It is a challenge for any
proposal that purports to explain the extreme homogeneity and isotropy
of the Universe to also explain the amplitude and distribution of
these perturbations.

\subsection{Inflation}

The inflationary universe idea was proposed by A. Guth to address these 
issues by relying on the dynamics of a phase transition. 
The horizon problem can be addressed if there existed an epoch
in the universe when the particle horizon was growing faster than 
the Hubble horizon. Later this phase ends and we return to radiation 
and matter dominated Universe. 

If the Universe was purely radiation dominated during its entire
early history particle horizon could not have grown faster than 
Hubble horizon. However, if there is an unusual equation of state 
obeyed by the source terms of Einstein's equations then this is 
possible. Our study of first order phase transitions suggests a
possible scenario. We have seen that there is a possibility for
the system ( the universe) to be trapped in a false vacuum. Exit
from such a vacuum occurs by quantum tunneling. If this tunneling 
rate is very small, the vacuum energy of the false vacuum will
dominate the energy density of the Universe. Vacuum energy of a
scalar field has just the right property to ensure rapid growth of
particle horizon, keeping the hubble horizon a constant.
If inflation occurs, the flatness problem also gets automatically
addressed. 

Consider the energy momentum tensor of a real scalar field 
\bearrs
T^\mu_\nu&=& \p_\nu \phi \p^\mu \phi - \delta^\mu_\nu\mathcal{L}\\[10pt]
&=& \p_\mu \phi \p^\mu \phi - \delta^\mu_\nu
\l( \f{1}{2} \p_\la \phi \p^\la \phi - V (\phi ) \r)
\eearrs
If the field is trapped in a false vacuum, its expectation value is
homogeneous over all space and is also constant in time. This
means all the derivative terms vanish and 
$ T^\mu_\nu \rar  V_0 \delta^\mu_\nu $ where $V_0$ is the value of the
potential $V$ in the false vacuum, called the vacuum energy density. 
Now for an isotropic and homogeneous fluid,
the energy-momentum tensor assumes a special form, 
$\textrm{diag}(\rho, -p, -p, -p)$. That is, the off diagonal terms are zero, 
the three space dimensions are equivalent, and the non-zero entries have
the interpretation of being the usual quantities $\rho$ the energy density
and $p$ the pressure. Thus comparing this form with that assumed
by the scalar field in a false vacuum, we see that the expectation value 
of the scalar field behaves like a fluid obeying the unusual equation of
state $p=-\rho$. 
Now the Friedmann equation becomes 
\[
\l(
\f{\dot{R}}{R}\r)^2 = \f{8\pi}{3} G\rho = \f{8\pi}{3} GV_0
\]
whose solution is 
$R(t) = R\l( t_i \r) \exp \l( H \l( t - t_i \r)\r)$, with $H^2=
(8\pi/3)V_0$ and $t_i$ is some initial time.

\subsection{Resolution of Problems}
\subsubsection{Horizon and flatness problems}
During the inflationary epoch we assume an exponential expansion and
a constant value of Hubble parameter $H$ (defined without special
subscript or superscript since $H_0$ is reserved for the current value of
$H$). Now consider the particle horizon, or equivalently, the luminosity
distance at any epoch $t$
\[
d_H = e^{Ht} {\dis \int}^t_{t_i} \f{dt^\pr}{e^{Ht^\pr}} \approx
\f{1}{H} e^{H\l( t-t_i\r)}
\]

\begin{figure}[hbp]
{\par \centering \resizebox*{0.8\textwidth}{!}
{\rotatebox{0}{\includegraphics{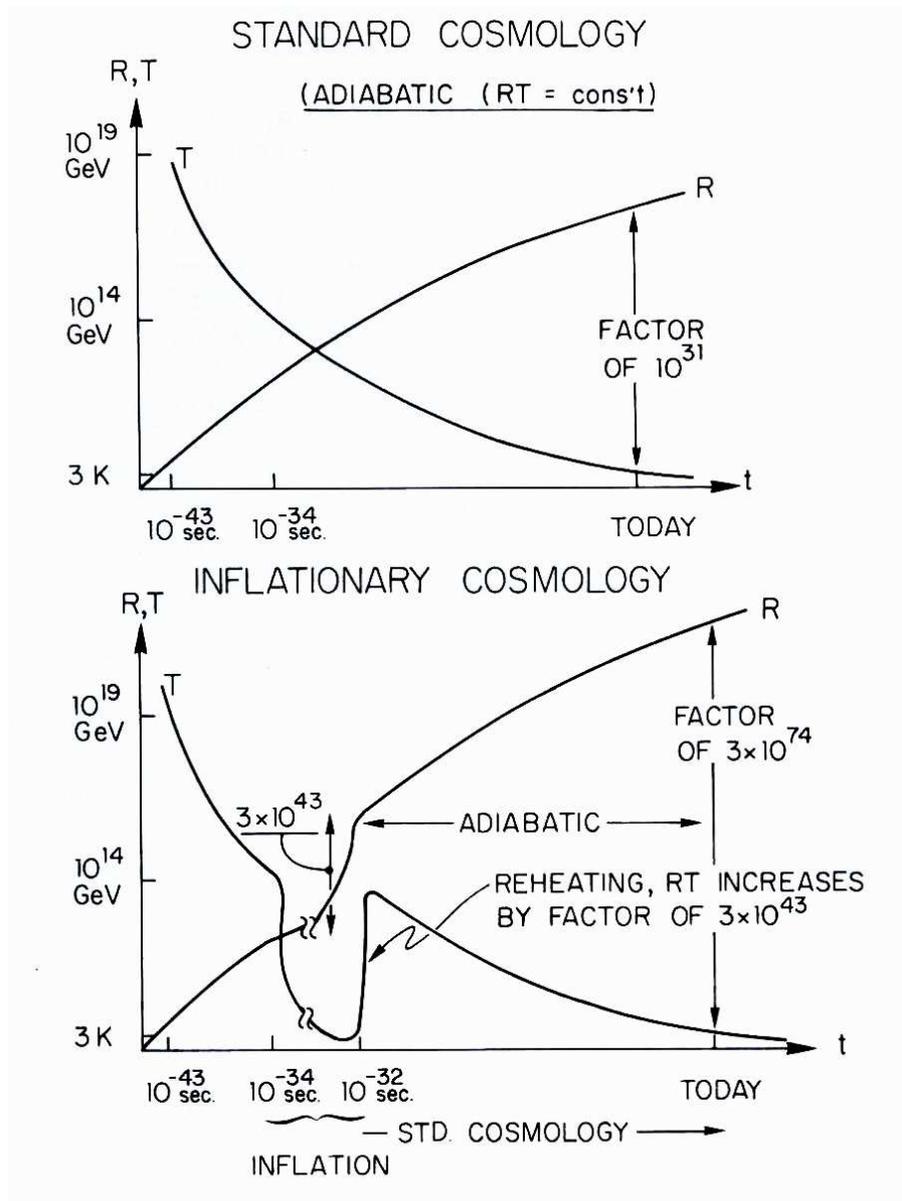}}} \par}\vspace{0.3cm} 
\caption{Comparison of simple FRW cosmology with inflationary cosmology. 
Figure courtesy Kolb and Turner.}
  \label{fig:stdvsinf}
\end{figure}

Thus the distance over which causal effects could be exchanged is
exponentially larger than the simple estimate $3 t_{dec}$ we used while
discussing the horizon problem. Suppose inflation lasted for a duration
$(t-t_i)\equiv \tau$. We can estimate what value $H\tau$ we need during 
inflation so that we are not seeing a large number of primordial horizon 
volumes but only about one. Suppose inflation ended leaving the Universe at
a temperature $T_{r}$ ( subscript ``r'' signifies reheat as explained
in next subsection). Then the $d_H$ of above equation rescaled to
today assuming radiation dominated Universe\footnote{The Universe has of 
course not been radiation dominated through out. But replacing 
later history of the Universe by matter dominated evolution
will not significantly alter these estimates since
the expansion is changed by a small change in the power law, $t^{2/3}$
instead of $t^{1/2}$. Assumption of radiation dominated expansion allows 
relating temperatures at two different epochs, as a good approximation.}  
should give current inverse horizon $H_0^{-1}$. Thus
\[
\l( \f{1}{H} e^{H\tau} \r) \times \l( \f{T_r}{T_0} \r)
\approx \f{1}{H_0}
\]

In above formula let us estimate $H/H_0$ again assuming radiation
dominated evolution since the $T_r$ till now,
\[
\f{H}{H_0} \sim \left. \f{t_0}{t}\right|_{FRW} \sim \l( \f{T_r}{T_0}
\r)^2 \sim \l( \f{10^{14} \mbox{ GeV}}{10^{-4} \mbox{ eV}}\r)^2 \sim
\l( 10^{27} \r)^2
\]
recall that $H^2\propto T^4$ and we inserted a Grand Unification scale 
$10^{14}$GeV as a possibility for $T_r$. Then $e^{H\tau}$ has to be $10^{27}$. 

We need an improvement upon this estimate. As we shall later see, 
there are theoretical reasons to believe that
$T_{r}\lesssim 10^9$GeV. In this case it is also assumed that there is
additional ( non-inflationary or mildly inflationary) stretching by
a factor $10^9$.  In this case we delink the reheat temperature $T_{r}$
from the vacuum energy density causing inflation, and assume the
latter to continue to be at GUT scale. In this case, 
\[
e^{H\tau} \sim \l( \frac{H}{H_0}\r)\times\l( \frac{t_0}{T_{r}}\r)
\times 10^{-9}\sim10^{18}
\]
where we have used estimate of the vacuum energy density 
$(0.03\mbox{eV})^4$ as 
derived from the directly observed value of current Hubble constant.
In the literature one often sees this estimate made with GUT scale taken to be 
$10^{16}$GeV so that the required value of $e^{H\tau} \gtrsim 10^{22} \sim e^{55}$.\\

Let us see the requirement to solve the flatness problem. We need to show that 
the term $|k|/R^2 $ becomes insignificant regardless of its value before inflation.
For naturalness we assume it to be comparable to $H^2$ as expected from Friedmann
equation. Now including scaling from inflation, the stretching by factor $10^9$
as introduced above, and subsequently during the era after reheating, we find
\[
\l( \f{k}{R^2} \r)_0 \sim \l( \f{k}{R^2}\r)_{\mbox{pre-inf}} \times
e^{-2\tau} \times \l( 10^9\r)^{-2} \l( \f{T_0}{T_r} \r)^2
\]
The left hand side is $(1-\Omega)H_0^2 $ which is insignificant.
The right hand side is the same small factor we estimated above, squared.
So this means the left hand side is reduced to a value $10^{-36}$ or $10^{-44}$
depending on the value of Grand Unification scale we take.

\subsubsection{Avoidance of unwanted relics}
A byproduct of Inflation is that it would also explain absence of exotic 
relics from the early Universe.
Typically a grand unified theory permits occurance of topological
defects such as cosmic strings or monopoles. They signifiy unusual local
vacua of spontaneously broken gauge theories which cannot evolve by 
unitary quantum mechanical processes to the simple vacuum.
Unlike normal heavy particle states, therefore, such defects cannot decay.

The natural abundance of such relics can be calculated by understanding
the dynamics of their formation.
Typically these events are the phase transitions characterised by specific
tempratures. If the naturally suggested adundances of these objects
really occured, they would quickly dominate the energy density of the
Universe, with possible exception of cosmic strings. This would be
completely contradictary to the observations.
On the other hand if the scale of Inflation was below the temperature
of such phase transitions, the density of topological objects formed would
be diluted by the large factor by which volumes expand during inflation.

Another class of exotic relics are the so called moduli fields, scalar
excitations arising in supersymmetric theories and String Theory.
They are generic because of the powerful symmtry restrictions on
potential energy functions in such theories. Inflation provides a solution
for some class of models for this case also.

\subsubsection{Resolution for density perturbations}
Finally the density perturbations are neatly explained by inflation. 
The scalar field is assumed to be in a 
semi-classical state. However quantum fluctuations do exist 
and these should in principle be observable.  
Inflationary era is characterized by Hubble parameter $H$ remaining
a constant while the scale factor grows exponentially. Thus the
wavelengths of various Fourier components of the fluctuations are
growing rapidly, leaving the horizon.

We shall take up in greater detail the theory of small perturbations 
in an expanding Universe in section \ref{sec:perturbations}. 
There we show that 
the amplitudes of the Fourier  modes of these perturbations remain 
frozen at the value with which they left the Hubble horizon. 
Eventually when the Universe becomes radiation dominated and subsequently
matter dominated, Hubble horizon $H^{-1}$ begins to grow faster than 
the scale factor and the wavelengths of the modes begin to become 
smaller than Hubble horizon.
This is the same reason as the solution of the horizon problem 
wherein apparently uncorrelated regions of distant space 
now seem to be correlated. They all emerged from the same
causally connected region and got pushed out of the horizon during the
inflation epoch.

The result of this evolution of the fluctuations is that when they
re-enter the horizon they all have the same amplitude. These fluctuations
then influence the rest of the radiation and matter causing fluctuations 
of similar magnitude in them. This is the explanation for the 
scale invariant matter density perturbations represented by 
distribution of galaxies. We shall take up the details of fluctuations
in the next section.

\subsection{Inflaton Dynamics}
Inflation is a paradigm, a broad framework of expectations rather than
a specific theory. The expectations can be shown to be fulfilled if 
a scalar field dubbed ``inflaton'' obeying appropriate properties exists. 
If inflation is implemented by such a scalar field, we 
need to make definite requirements on its evolution. We assume that 
its evolution leads the Universe through the following three phases
\begin{itemize}
\item Inflationary phase
\item Coherent oscillation phase
\item Decay and re-heating phase (``re-heat'' only for low field scenario)
\end{itemize}

\begin{figure}[htbp]
{\par 
\centering
\resizebox*{0.4\textwidth}{!}{\rotatebox{0}{\includegraphics{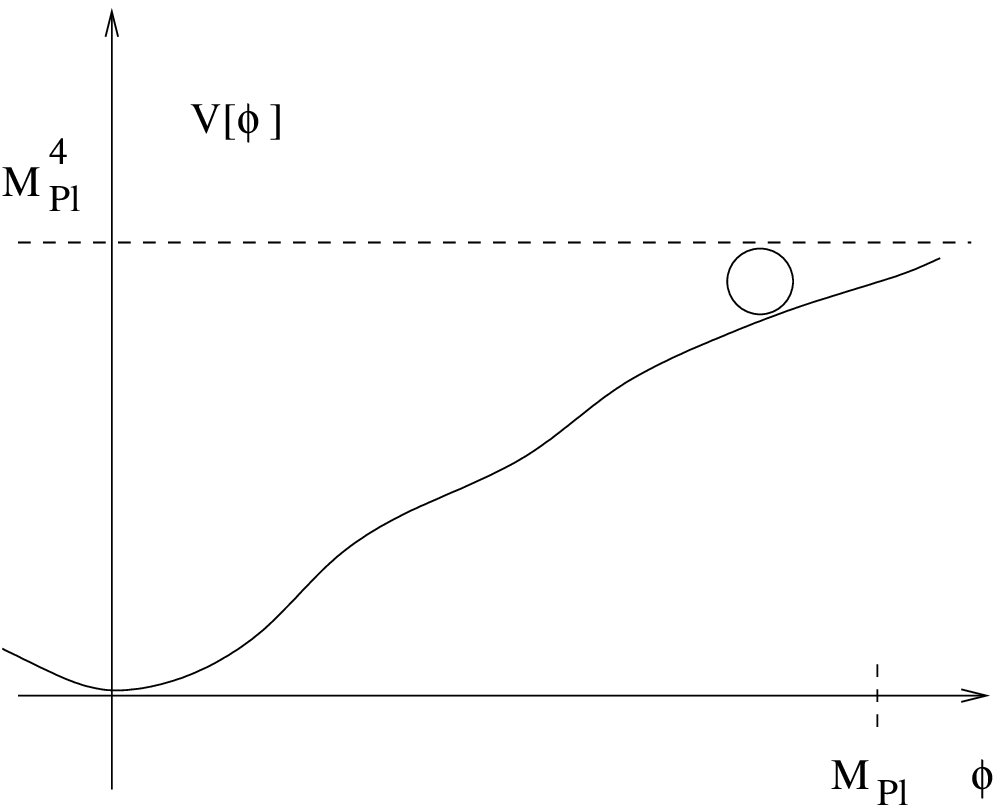}}}
\hspace{5mm}
\resizebox*{0.4\textwidth}{!}{\rotatebox{0}{\includegraphics{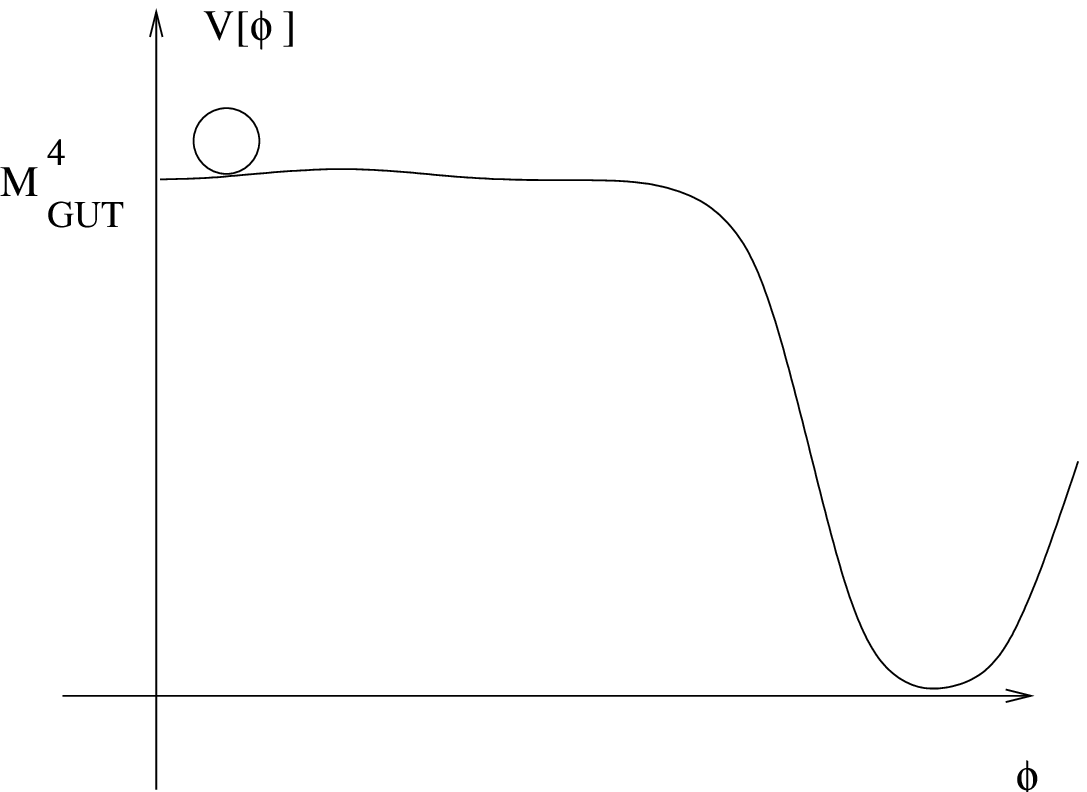}}}
\par}\vspace{0.3cm} 
\caption{Sketches of stipulated effective potentials with the corresponding 
initial value of the inflaton field indicated by a small circle 
( a ``ball'' ready to ``roll down'' the shown profiles). Left panel shows 
the High Field scenario, the right panel shows the Low Field scenario}
  \label{fig:inflationscenarios}
\end{figure}

Of these three phases, the inflationary phase addresses the broadest 
requirements discussed in previous subsection. To obtain exponential
expansion we need constant vacuum energy. This means that the field
$phi$ has a value where $V(\phi)\neq 0$ and also that $\phi$ continues
to remain at such a value. The simplest such possibility is a false
vacuum, a local minimum of the effective potential which is not a
global minimum. But this is problematic because this kind of state
can be far too stable and the exit from it keeping the Universe 
permanently inflating. Two possibilities which are strong candidates due to 
phenomenological reasons are the so called High Field (older name Chaotic 
Inflation) or the Low Field ( older name New Inflation) scenarios of 
inflation. These are shown in figure \ref{fig:inflationscenarios}. 
In the High Field case the initial value of the field is close to 
Planck scale and no clear barrier separating it from the low energy 
true minimum. However its dynamics governed by Planck scale effects 
is ''chaotic'' and keeps it at a very high energy for a long time. 
In the Low Field case,  the dynamics is usual field theory but 
the effective potential  function has a long plateau of very 
small slope. Assuming the initial value of the field at the top of
the plateau, this allows 
the field to sustain a position of large vacuum energy for a long time. 
In both cases, the field eventually moves towards
the low energy true minimum, via the next two phases. 
A third possibility  which is appealing for supersymmetric unified theories
is called the Hybrid scenario. It involves two fields, one which keeps
the Universe initially at a high field value, and the second field which
becomes more dominant at a later stage, causing a rapid roll down and 
exit from the high energy plateau. It will not be possible for us to 
discuss these scenarios in any detail in these notes.

Next, it is easier to explain the third listed phase, since it is necessary 
that the end point of inflationary expansion is a hot Universe.  Big Bang 
Nucleosynthesis is very  successful in explaining the natural 
abundance of elements. We need to assume a hot Universe of at least
a few MeV temperature for BBN to remain viable. 

Finally, the intermediate phase dominated by coherent\footnote{i.e., 
a special state in which it is possible to treat the field in the leading
order as if it were a classical field.} 
oscillations of the scalar field 
is almost certain to occur as inflation ends. This is because during
inflation the field is already in a coherent state, one in which it has 
homogeneous ( position independent ) value. This phase ends with
creation of quanta. In a large class of models this phase may be
of no particular interest. But in special cases when the coupling
of the inflaton to other matter is tuned to certain values, it
can lead to a variety of interesting effects which can have
observable consequences. One such possibility is a long duration
of coherent oscillations, which can be shown to mimic a Universe
filled with pressureless dust. Alternatively there can be particles
with special values of mass which can be shown to be produced 
copiously during the coherent oscillation epoch. Such effects 
are called ``preheating''  because heating of the Universe is
achieved not directly through the inflaton but by other particles
which have efficiently carried away the energy of the inflaton. 
Such phases can 
enhance the expansion achieved during the inflationary phase, 
the additional $10^9$ factor used in estimates in previous 
subsection. 

\subsubsection{Predictions and observables}
For a quantitative study we consider a scalar field $\phi$ with lagrangian
\[
S[\phi ] = {\dis \int} d^4x \sqrt{-g} \l( \f{1}{2} \p_\mu \phi
\p^\mu \phi - V (\phi ) + \mathcal{L}_{\phi-\mbox{matter}} \r)
\]
To discuss the inflationary phase we do not need 
$\mathcal{L}_{\phi-\mbox{matter}}$
containing the detail of the coupling of the scalar field to the rest 
of matter. Varying this action after putting the Friedmann metric
for $g_{\mu\nu}$ gives an equation of motion for $\phi$,
\[
\ddot{\phi} + 3H\dot{\phi} + V^\pr (\phi ) = 0
\]
The inflationary phase is characterized by a homogeneous value of
$\phi$ and a very slow time evolution so that there is domination
by vacuum energy. Mathematically one demands
\[
\ddot{\phi} \ll 3H\dot{\phi} 
\]
in other words we assume that the time scale of variation of the 
field $\phi$ encoded in $\dot\phi/\ddots\phi$ is small compared 
to the time scale $ H^{-1}$ of the expansion of the Universe.
Further, the assumption that the time scale of variation of $\phi$ 
is very small amounts to assuming that in the action we must have
\[
\dot{\phi}^2 \ll V_0 
\]
Thus while obtaining the Euler-Lagrange equations the $V$ term
continues to be important, but the higher time derivative of
$\phi$ can be dropped. So the evolution equation reduces to
\[
3H\dot{\phi} = - V^\pr (\phi )
\]
To ensure consistency of the above two conditions, divide  the 
simplified evolution equation by $3H$ and take a time derivative.
Recall that $H^2=(8\pi/3)G V(\phi) $. Thus $H$ can be implicitly
differentiated with respect to time as shown in the equation below
\[
\ddot{\phi} = - \f{V^"(\phi )\dot{\phi}}{3H(\phi )} + \f{V^\pr
(\phi )}{3H^2(\phi )} H^\pr (\phi ) \dot{\phi}
\]
We now write the equation in three equivalent forms
\bearrs
\f{\ddot{\phi}}{3H\dot{\phi}} &=& - \f{V^"(\phi
)}{9H^2} + \f{V^\pr}{9H^3} H^\pr\\[10pt]
&=& - \f{V^" (\phi )}{9\times \f{8\pi}{3} GV_0} + \f{V^\pr \l(
\f{4\pi}{3} \r) GV^\pr}{9\times \l( \f{8\pi}{3} G \r)^2
V^2}\\[10pt]
&=& - \f{V^"}{3\times 8\pi G V_0} + \f{1}{2} \f{1}{3\times 8\pi G}
\l( \f{V^\pr}{V} \r)^2
\eearrs
where use has been made of the preceding assumptions as also the
relation $2HH^\pr= \f{8\pi}{3} GV^\pr$.
 
Inspecting the above equations we define three parameters
\[
\epsilon \equiv \f{Mp^2}{16\pi} \l( \f{V^\pr (\phi )}{V(\phi
)}\r)^2, \;\;\; \eta \equiv \f{Mp^2}{8\pi} \f{V^"(\phi )}{V(\phi
)}
\]
and 
\[
\xi \equiv \f{\ddot{\phi}}{H\dot{\phi}} = \epsilon + \eta
\]

The requirements of the inflationary phase viz., large vacuum
energy and a vary slow roll towards the true minimum mean that
\[
\epsilon \ll 1, \;\; \mid \eta \mid \ll 1 \; \mbox{ and } \xi \sim
0 ( \epsilon , \eta )
\]
We use these in the criterion that if either of these quantities 
becomes large, inflationary phase ends. These have come to be called
the ``slow roll'' parameters characterizing inflation.

For several decades inflation remained a theoretical paradigm.
However with precision cosmological experiments such
as the Hubble Space Telescope (HST) and the Wilkinson Microwave
Anisotropy Probe (WMAP) yielding valuable data we face the exciting 
prospects of verifying and refining the paradigm, and also deducing the
details of the dynamics of the inflaton field. It is possible to 
set up a relationship between slow roll parameters introduced
above and the temperature fluctuation data of the microwave
background radiation. Similarly the large scale galaxy surveys
such as 2dF GRS ( 2 degree Field Galaxy Redshift Survey) and 
6dF GRS provide detailed data on distribution of galaxies which
can be counter checked over a certain range of wavenumbers against the 
fluctuations as observed in WMAP and also against the specific 
dynamics of the inflaton.

Finally let us work out a simple example of how we can relate intrinsic
properties of the effective potential to observable features of
inflation. We can for instance compute the number of e-foldings $N$
in the course of $\phi$ evolving from initial value $\phi_i$ to a 
value $\phi_f$, given the form of the potential.
\bearrs
N \l( \phi_i \rar \phi_f \r)&\equiv& \ln \l( \f{R\l(
t_f \r)}{R \l( t_i \r)} \r) = {\dis \int}^{t_f}_{t_i} Hdt = {\dis
\int} \f{\dis H}{\dis \dot{\phi}} d\phi\\[10pt]
&=& - {\dis \int}^{\phi_f}_{\phi_i} \f{\dis 3H^2d\phi}{\dis V^\pr}\\
\eearrs
where we assumed $t_f \sim H^{-1} \l( \phi_f \r)$. Now shift the global minimum of the effective potential to be  
at $\phi = 0$ so $\phi_i \gg \phi_f$
\bearrs
&=& - 8\pi G {\dis \int}^{\phi_f}_{\phi_i} \f{V(\phi )}{V^\pr
(\phi )} d\phi 
\eearrs
This formula can be used to relate the dominant power law in
the effective potential with the number of e-foldings. For
$V(\phi ) = \la \phi^\mu$,
\bearrs 
N \l( \phi_i \rar \phi_f \r) &=& \f{4\pi}{\nu} G \l( \phi_0^2 -
 \phi^2_f \r)\\[10pt]
 &\approx& \l( \f{4\pi}{\nu} G \r) \phi^2_{int}
\eearrs

\section{Density perturbations and galaxy formation}
\label{sec:perturbations}
An outstanding problem facing FRW cosmological models is formation 
of galaxies. If the Universe emerged from the Planck era perfectly
homogeneous and isotropic, how did the primordial clumping of
neutral Hydrogen occur? Without such clumping formation of
galaxies and in turn stars would be impossible.

The related observational facts are also challenging. The fluctuations
in the average density have resulted in a distribution of galaxies and
clusters of galaxies. What is remarkable is that these fluctuations 
exist at all observable scales. Further, the observed fluctuations 
seem to have originated from seed fluctuations which were of the
same magnitude, approximately one part in $10^{5}$,  independent 
of the scale at which we study the fluctuations. This statement of
scale invariance began as a hypothesis, known  as the 
Harrison-Zel'dovich spectrum but has been remarkably close to 
the extensive experimental evidence accumulated over the last fifty years.
The main sources of current data are Sloan Digital Sky Survey (SDSS),
Two degree field Galaxy Redshift Survey (2dF GRS), quasar redshift surveys, 
"Lyman alpha forest" data etc, collectively called Large Scale Structure 
(LSS) data.

Here we shall present a brief overview of the formalism used for
studying fluctuations. We shall also show that
the inflationary Universe is in principle a solution, providing
scale invariant fluctuations. The magnitude of resulting fluctuations 
however is too large unless we fine tune a parameter to required value.

\subsection{Jeans Analysis for Adiabatic Perturbations}
First we study the evolution of perturbations in a non-relativistic fluid.
We study the continuity equation, the force equation and the equation
for gravitational potential assuming that the state of the fluid provides
a solution a solution to these. We then work out the equations satisfied 
by the perturbations. Fourier analysing the perturbations, it is found 
that modes with wavelengths larger than  a critical value of the 
wavelength the perturbations  are not stable. 
 
The relevant equations for the mass density $\rho(\mathbf{x},t)$,
velocity field $ \mathbf{v}(\mathbf{x},t)$ and the newtonian gravitational
potential $\phi(\mathbf{x},t)$ are
\bearrs
\f{\p \rho}{\p t} + \mathbf{\nabla} \cdot \l( \rho \mathbf{v} \r) &=&
 0\\[10pt]
\f{\p \mathbf{v}}{\p t} + \l( \mathbf{v} \cdot \mathbf{\nabla} \r)
\mathbf{v} + \f{1}{\rho} \mathbf{\nabla} p + \mathbf{\nabla} \phi &=& 0\\[10pt]
\nabla^2\phi &=& 4\pi G\rho
\eearrs
We now split the quantities into average values $\bar\rho$, $\bar p$ 
and space-time dependent perturbations $\rho_1$, $p_1$
\[
\rho(\mathbf{x},t) = \bar\rho + \rho_1(\mathbf{x},t)  \qquad  
p(\mathbf{x},t) = \bar p + p_1(\mathbf{x},t)
\]
and similarly for velocity. However, for a homogeneous fluid, average
velocity is zero so $\mathbf{v} $ and $\mathbf{v}_1 $ are the same. 
It is reasonable to assume that there is no spatial variation in equation 
of state. This means the speed of sound is given by
 \[
v^2_s = \l( \f{\p p}{\p \rho}\r)_{\mbox{adiabatic}} =
 \f{p_1}{\rho_1}
 \]
Thus the equations satisfied by the fluctuations are,
\bearrs 
\f{\p \rho_1}{\p t} + \bar\rho \mathbf{\nabla} \cdot
\mathbf{v}_1 &=& 0\\[10pt]
\f{\p \mathbf{v}_1}{\p t} + \f{v^2_s}{\bar \rho} \mathbf{\nabla}\rho_1 +
\mathbf{\nabla}\phi_1&
=&0\\[10pt]
\nabla^2\phi_1&=& 4\pi G\rho_1 
\eearrs
From these coupled equations we obtain a wave equation for $\rho_1$,
\[
\f{\p^2\rho_1}{\p t^2} - v^2_s \nabla^2\rho_1 = 4\pi G \bar\rho \rho_1
\]
whose solution is
\bearrs
 \rho_1 \l( \mathbf{r}, t \r) &=&  Ae^{\l( i\omega t - i
\mathbf{k} \cdot \mathbf{r} \r)}\\[10pt]
\mbox{with}\; \omega^2&=& v^2_s k^2 - 4\pi G \bar\rho 
\eearrs
The expression for $\omega $ suggests the definition of a critical wavenumber, 
Jeans wavenumber, $ k_J \equiv \l( 4\pi G\bar\rho/v^2_s \r)^{1/2}$.
For $k \ll k_J$ we get exponential growth, i.e., instability.
We can understand this result by associating with a wavelength $\la$
a hydrodynamic timescale $\tau_{\mbox{hyd}}$ $\sim\la/v_s \sim 1/kv_s$.
This is the timescale during which pressure differences will be
communicated by perturbations with wavelength $\la$. Next we associate
the timescale $\tau_{\mbox{grav}}=\l( 4\pi G \rho_o \r)^{-1/2}$ with 
the gravitational influences. The result above says that for the
wavelengths for which hydrodynamic response is slower to propagate
than gravitational influences, the latter win and cause 
a gravitational collapse. 

We can now define Jeans mass 
\[
M_J \equiv \f{4\pi}{3} \l( \f{\pi}{k_J}\r)^3 \rho_o =
\f{\pi^{5/2}}{6} \f{{v_s}^3}{G^{3/2} \rho_0^{1/2}}
\]
We can deduce that a homogeneous mass bigger than this value is 
susceptible to gravitational collapse.

\subsection{Jeans Analysis in Expanding  Universe}
We need to extend the above treatment to the case of expanding Universe.
Firstly we must let the mean density and pressure be time dependent due to
change in cosmological scale factor. For instance, the mean density in a matter
dominated universe will scale as $\bar\rho(t) = \bar\rho \l( t_{i} \r)
\l(R(t_{i}/R(t)\r)^{3}$ where $t_{i}$ denotes some ``initial'' reference time.

From now on we shall not follow the evolution of the other quantities but 
focus on the energy density.  We introduce the dimensionless quantity 
$\delta = \rho_1/\bar\rho$ from which the obvious $R(t)$ dependence gets scaled out. 
Further, using the FRW metric in the comoving form, we treat $\mathbf{r}$ to be
dimensionless and $t$ and $R(t)$ to have dimensions of length ( equivalently, time). 
Introduce the Fourier transform
\[
\delta(\mathbf{x},t) =
\int \f{d^3k}{(2\pi)^3} \delta_\mathbf{k}(t)
e^{\l( - i \mathbf{k} \cdot \mathbf{r} \r)}
\]
It can be shown that these Fourier modes obey the equation
\[
\ddot{\delta}_k + \f{2\dot{R}}{R} \dot{\delta}_k 
+ \l(\f{v^2_sk^2}{R^2} - 4\pi G\bar\rho(t) \r) \delta_k = 0 
\]
We see that the modified Jeans wave number defined by 
\[
k^2_J \equiv 4\pi G\bar\rho(t)R(t)^2/v^2_s
\]
plays a crucial role, in that at any given epoch $t$, wavelengths shorter 
than $\sim 1/k_J$ are oscillatory and hence stable. 

We shall now study the fate of the long wavelengths, the ones significant
on cosmological scales. We shall see that the unstable modes 
grow in time, but instead of exponential growth they can have power law growth.
We have assumed $k\rar 0$. Further, using FRW equation we can replace the $G\bar\rho(t)$
term by $(3/2)(\dot R/R)^2$ in a spatially flat universe. Then for matter dominated 
universe with $R\propto t^{2/3}$, we get 
\[
\ddot{\delta} + \f{4}{3t} \dot{\delta} - \f{2}{3t^2} \delta = 0
\]
The solutions are
\bearrs
\delta_+(t) &=& \delta_o \l( t_i \r) \l( \f{t}{t_i}\r)^{2/3}\\[10pt]
\delta_-(t) &=& \delta_- \l( t_i \r) \l( \f{t}{t_i} \r)^{-1} 
\eearrs

Studying the examples of other power law expansions of scale factor $R$, one may
conclude that the expansion of the universe keeps pulling apart the infalling matter 
and slows down the growth of Jeans instability. However
In the case of de Sitter universe it is found that exponential instability persists.
In this case the long wavelength modes obey the equation
\[
\ddot{\delta} + 2H\dot{\delta} - \f{3}{2} H^2\delta = 0
\]
so that substituting $\delta \sim e^{\al t}$ we find 
\[
\al^2 + 2H\al - \f{3}{2} H^2 =0
\]
This has the roots $\al_\pm = - H \pm \sqrt{H^2+3/2}$.
One root is negative definite signifying decaying exponential
though one positive root persists, $\al_+ = \sqrt{H^2+3/2} - H$.

\subsubsection{Fate of the super-horizon modes}
Let us now return to the idea of inflationary universe as the source of primordial
perturbations. The basic hypothesis is that the quantum mechanics of the
``inflation" scalar field causes fluctuations in its expectation value. 
These then manifest as perturbations in the classical quantity, the energy density.
This assumption is expressed as
\[
\rho_1\equiv \delta\rho = \f{\delta V}{\delta \phi}\delta \phi
\]
where 
\[
\delta \phi \equiv \langle (\phi - \langle \phi \rangle)^2 \rangle
\]
and the expectation values are computed in an appropriately chosen state. This
choice is not always easy. In general this is a static, translation invariant 
state with similar properties shared by the vaccum expectation values. In 
an expanding Universe the corresponding symmetries available are those of 
the space-time metric, namely the FRW metric. Since the time translation 
symmetry is lost, there are several conceptual issues.
Fortunately the de Sitter metric has a sufficiently large group of symmetries
permitting a fairly unique choice of the vaccum. Inflationary universe 
resembles the de Sitter solution over a substantial length of time so 
that we can adopt the answers obtained for the de Sitter case.

Decompose $\delta\phi$ into Fourier modes with the same conventions as
in the preceding section.
During the inflationary phase, the expectation value of $\phi$ remains approximately
constant. Hence by appropriate shifting of the field, dynamics of $\delta\phi$ and $\phi$
are the same. The equations of motion for the modes of $\delta\phi$ are then
\[
\delta \ddot{\phi}_k + 3H \delta \dot{\phi}_k + k^2 \f{\delta
\phi_k}{R^2} = 0
\]
Then for ``super-horizon" fluctuations with wavenumbers satisfying $k \ll RH$,
we can ignore the last term and the non-decaying solution is $\delta \phi_k$ = constant.
This is a crude argument to justify that the fluctuations at
this scale become constant in amplitude. But constant amplitude would mean
vanishing time derivatives, so the third term can't be smaller than the
first two. In order to consistently ignore the third term relative to the second, 
we need to additionally assume that if the time scale of variation of $\delta \phi_k$ is $\tau$, 
then $1/\tau > k/R$ in addition to $k \ll RH$. The two inequalities together imply
$1/\tau \gg H$. Thus the fluctuations are constant in the sense that the time scale 
of their variation is much less than the natural time scale of the geometric background, $H^{-1}$.

The above statement can be made more precise  using
Quantum Field Theory in curved spacetime, where it can be shown that
in de Sitter universe, for a massless scalar field, the fluctuations in
$\phi$, after appropriate cut-off procedure, are given by 
\[
\langle\phi^2\rangle = \l( \f{H}{2\pi}\r)^2
\]
The assumption in this calculation is that the state chosen is de Sitter
invariant. Thus variations that do not respect this invariance don't
constribute and we get the expected constant result. Further, since no new
scales are introduced through the choice of the state, the only dimensionful 
quantity available in the problem is $H$, which effectively determines the 
magnitude of the fluctuations.

We can now see the qualitative features which make inflation so appealing
for generating scale invariant perturbations.
For all normal kind of states of matter and energy, the scale
factor grows as power law $t^s$, with $s<1$. Thus $H^{-1}$ 
grows as $t$ and all wavelengths $\la_{phy}$ scale as 
$\l(\la_{phy}/R\l(t_1\r)\r) \times R(t)$ and hence keep falling
inside the horizon.  On the other hand, in the inflationary phase, 
$H^{-1}$ = constant and wavelengths grow exponentially, 
$\la_{phy}(t) \propto \la_i e^{H\l( t-t_i\r)}$ with some 
reference initial time $t_i$. Later, after inflation ends, 
$H^{-1}$ begins to grow faster than other length scales 
and steadily catches up with fluctuations of increasing values of wavelengths. 
Between leaving $H^{-1}$ and re-entering
$H(t)^{-1}$ later, the amplitude of the fluctuations remains frozen by
arguments of preceding para. Note that for other power law expansions
$t^s$ with $s>1$ also, the wavelengths grow faster than the Hubble
horizon. However, the amplitude of these wavelengths will not remain
constant and will not reproduce the scale invariant spectrum after
re-entering the horizon. Some of these arguments will become clearer
in the following subsection.

\subsubsection{Connection to density and temperature perturbations}
The problem of galaxy formation is to predict the observed pattern
of clumping of luminous matter. Also, since matter and radiation were
in equilibrium upto decoupling, the fluctuations in matter have also 
to be reflected in the fluctuations in the temperature of the CMBR. 

The fluctuation we are referring to are in the spatial distribution.
These are mathematically characterized by the auto-correlation
function
\[
(\Delta \rho)^2(\mathbf{r}) \equiv
\langle \delta \rho(\mathbf{x})\delta\rho(\mathbf{x}+\mathbf{r})\rangle
\]
where on the right hand side, an averaging process is understood.
For a homogeneous medium, the locations $ \mathbf{x}$ are all equivalent and
this dependence drops out at the end of averaging process.
Introducing the Fourier transform $\delta_\mathbf{k}$, we can show that
\[
(\Delta \rho)^2(\mathbf{r}) = \int dk \f{k^2}{2\pi^2}|
\delta_{\mathbf{k}} |^2 \f{\sin kr}{kr}
\]
Now the rms value $\Delta\rho_{rms}$ at a given point is the square-root of
this auto-correlation function for $r=0$.
Accordingly, taking the limit $kr\rar0$ in the above expression, 
we get 
\[
(\Delta \rho_{rms})^2 = \int \f{dk}{k} \mathcal{P}(k)
\]
with
\[
 \mathcal{P}(k)\equiv \f{k^3}{2\pi^2} |\delta_{\mathbf{k}}  |^2
\]
$\mathcal{P}$ represents
the variation in $(\Delta \rho_{rms})^2$ with variation in $\ln k$.
The aim of experiments is to determine the quantity $\mathcal{P}$.

Linear perturbation theory used above is valid only for small
fluctuations. Once a fluctuation grows in magnitude it begins
to be controlled by non-linear effects. We can estimate the intrinsic
scale upto which the mass of a typical galaxy could have been in 
the linear regime.  Using the value of the $\bar\rho$  to be the present
abundance of non-relativistic matter ( $\approx 10^{-29}$g/cc) and
bringing out a factor of $10^{11}$ solar masses, the mass in a
sphere of diameter of wavelength $\la$ is
\[
M \simeq 1.5\times 10^{11} M_\odot (\Omega_0 h^2) \l(\f{\la}{\textrm{Mpc}}\r)^3 
\]
Assuming $10^{12}$ solar masses per galaxy, this gives the size $\la$
to be $1.9$ Mpc, far greater than the actual galactic size $30$ kpc.
The $\la$ found here represents the size this mass perturbation would
have had today had it not entered the non-linear regime.

Present data are not adequate to determine the spectrum $\mathcal{P}$
over all scales. However too large a magnitude of fluctuations  at
horizon scale would have been imprited on temparature fluctuations
of CMB, which it is not. Likewise large fluctuations at smaller
scales could have seeded gravitational collapse and given rise to a large
number of primordial black holes, which also does not seem to be 
the case. Hence the spectrum must not be varying too greatly over
the entire range of wavenumbers. It is customary to assume the
spectrum of $|\delta_{\mathbf{k}} |^2$ to not involve any special
scale, which means it must be a power law $k^n$. Further, a fluctuation
of physical scale $\la$ contains mass $M\sim \la^3 \sim k^{-3}$.
Hence the spectrum $\mathcal{P}$ $\sim M^{-1-n/3}$. Now if we
make the hypothesis that the spectrum of perturbations seems to be
independent of the scale at which we observe it, we expect
$\mathcal{P}\sim$ a constant, i.e. independent of $M$. For this to be 
true, $n$ must be $-3$. 

In the analyses of WMAP data of CMB it is customary to normalize the
rms perturbation spectrum by its observed value at $(8/h)\sim 11$Mpc and 
denote it $\sigma_8$. WMAP reports the best fit value to be $0.8$. Further,
$(\Delta \rho_{rms})^2$ is parameterized as $k^{(-1+n_s)}$ where the 
subscript in $n_s$ signifies scalar perturbations. A very long epoch 
of perfectly  de Sitter inflation
would produce $n_s=1$ and a perfectly scale invariant spectrum.   
Given a specific model of inflation the small departures of $n_s$ 
from unity can be calculated as a function of $k$.
This mild dependence of $n_s$ on $k$ is referred to as "running of the index" 
of the power law. Current WMAP data ( ie horizon scale perturbations)
seem to suggest $n_s=0.97$ but the direct observations of LSS data 
( galaxies and clusters of galaxies) suggest $n_s>1$.  

Likewise a formalism exists for relating the temperature fluctuations
with the density perturbations and in turn with the scalar field fluctuations.
We shall not go into it here and the reader is referred to the references.

\subsection{Density Fluctuations from Inflation}
We now show how the inflation paradigm along with the knowledge of
the form of the scalar potential helps us determine the magnitude
of the scalar field fluctuations. 

Suppose we wish to know the fluctuation in a scale of size of
our present horizon. According to the derivation in previous
subsection we need to know the value of the perturbation
when it left the horizon in the inflationary era. And we need
to know the number of e-foldings the inflationary universe went through
before becoming radiation dominated. It is the latter fact which
then determines the later epoch when the same scale re-enters the horizon.

Let us trace a physical scale $\ell_0$ today by keeping track of
corresponding co-moving value $\ell$.
We have to consider the evolution in two parts. From the present we
can only extrapolate back to the time when the current hot phase of 
the Universe began, i.e. the ``re"heated\footnote{We remind the reader that
it is possible in some inflationary scenarios for the Universe to never 
have been in thermal equilibrium before this stage. Hence the prefix 
``re" is purely conjectural though conventional.} phase.
Prior to that was the phase of inflaton oscillation and decay.
Reheating is assumed to be complete at a time $t_d$, the decay
lifetime of the inflaton or its product particles.

Thus the size of a scale $\ell$ can be extrapolated to the epoch
$t_f$ when inflation ended, (i.e., the slowness conditions on the evolution 
of the scalar field ceased to be valid) by
\[
\ell_f = \ell \l( \f{T_o}{T_r} \r) \l( \f{R\l( t_f
\r)}{R\l(t_d\r)}\r)
\]
The last ratio can be estimated if we are given the effective potential
$V(\phi)= \la \phi^\nu$ and a formalism for the dissipation of the
inflaton vacuum energy. We shall not pursue these details here but
claim that this can be calculated to be
\bearr 
\f{R\l( t_f \r)} {R\l( t_d\r)}
&=& \l( \f{t_f}{t_d}\r)^{(\nu + 2)/3\nu}\\[10pt]
&=& \l( \f{t_f}{t_d} \r)^{1/2}\qquad  \mbox{for} \;\nu=4 
\eearr

Now $\ell_{phys} \l( t_f \r) = H^{-1} \; e^{N_\ell}$  where
$N_\ell$ is the number of e-foldings between the time the specific scale
attained the value $H^{-1}$, (i.e. became comparable to the horizon) and 
the end of inflation.  
Substituting the current value of the horizon in the above expressions 
finally gives $N_{H_0^{-1}}\approx 50 - 60$.

We now trace the magnitude of the perturbation through this
exit from horizon followed by the re-entry at present epoch.
It turns out that $\delta\rho/\rho$ is a physically ambiguous quantity to
follow through such an evolution. This is because choice of a
particular time coordinate amounts a choice of a gauge in General Relativity.
The gauge invariant quantity to focus on has been shown to be $\zeta = \delta
\rho/(p+\rho)$. 

We seek the value of the numerator at a late epoch when inflation has ended.
The denominator at this epoch is determined by $p=0$ and the energy density 
which is dominated by the kinetic term. The value of $\zeta$ at the inflationary
epoch is known from preceding arguments about perturbations on scales comparable
to horizon. Here
\[\delta \rho= \f{\delta V}{\delta \phi}\delta\phi= V'(\phi)\f{H}{2\pi}
\]
where the $\delta\phi$ is estimated from QFT calculation of the rms value.
Further, we replace ${\dot\phi}^2$ by using the slow roll condition of
inflation, $3H{\dot\phi}=-V'(\phi)$. Thus we equate
\bearrs
\l(\f{\delta\rho}{\dot{\phi}^2}\r)_{\ell \sim H^{-1}} 
&=& \left. \zeta\right|_{\ell \sim H^{-1}} \\ [10pt]
&=&\left.\f{V^\pr (\phi ) H(\phi )}{2\pi \dot{\phi}^2}\right|_{\ell \sim
H^{-1}}\\ [10pt]
&=& \l( \f{9H^3(\phi )}{2\pi \l( V^\pr (\phi
)\r)}\r)_{\ell \sim H^{-1}}
\eearrs
Thus in matter dominated era when $p=0$, we have recovered
\[
\l( \f{\delta \rho}{\rho}\r)_\ell = \l( \f{2}{5} \r)
\times 8\sqrt{6\pi} \f{V^{3/2} \l( \phi_\ell \r)}{M^2_P V^\pr \l(
\phi_\ell \r)}
\]
where we reexpress $G\equiv 1/M^2_P$, the squared inverse of the Planck mass in 
natural units. The $2/5$ factor is acquired during transition from radiation dominated
to matter dominated era.

We thus need the values of $V$ and $V'$ at the value $\phi_\ell$. We do
not really know the latter directly. But we can determine it if we know the
number of e-foldings between its crossing the horizon and the end of inflation.
Inverting the relation
\[
H^{-1} \l( \phi_\ell \r) = \ell_{phy} \l( t_f \r) e^{N_\ell}
\]
for $\phi_\ell$ and also using
\[
N_\ell \l( \phi_\ell \rar \phi_f \r) = \int Hdt = {\dis
\int}^{\phi_f}_{\phi_\ell} \f{H}{\dot{\phi}} d\phi \rar \pi G
\phi^2_\ell
\]
where the last arrow gives the answer corresponding to the form $\la\phi^4$ 
for the effective potential and is obtained by consistently using the slowness 
condition and  the FRW equation.
This gives us the number of e-foldings between horizon crossing by this scale and
the end of inflation. Therefore, trading $N_\ell$ for $\phi_\ell$ we get 
\[
\l( \f{\delta\rho}{\rho}\r)_\ell = \f{4\sqrt{6\pi}}{5} \la^{1/2}
\l( \f{\phi_\ell}{M_P} \r)^3 = \f{4}{5} \sqrt{6\pi} \la^{1/2} \l(
\f{N_\ell}{\pi}\r)^{3/2}
\]
We have arrived at a remarkable mathematical relationship, expressing
the magnitude of perturbations visible in the sky today with the 
parameters of the effective potential that drove the primordial inflation.
We can take the fluctuations $\delta\rho/\rho$ to be as visible in 
the CMB temperature fluctuations, $\delta T/T \sim
6\times10^{-5}$. Assuming  $N = 55$ as needed for solving the horizon and
flatness problems, we find that we need the value of $\la \sim 6\times 10^{14}$.
This is a tremendous theoretical achievement. Unfortunately
the required numerical value is unnaturally small and it appears that
we have to trade the fine tuning required to explain the state of the
Universe with a fine tuning of a microscopic effective potential.

\section{Relics of the Big Bang}
As the Universe cools reaction rates of various physical processes
become slow. When they become slower than the expansion rate of
the Universe, the entities governed by those reactions no longer
interact and remain as residual relics. The mathematical description 
for these events is provided by Boltzmann equations which can be used
to infer the relative abundance of these relic particles which can
be in principle observed today.
The term relic applies to a wide variety of objects, including extended
objects like cosmic strings or domain walls but we shall be dealing only
with particle like relics in these notes.

\subsection{Boltzmann Equations}
Boltzmann equations describe the approach to equilibrium of a system
that is close to equilibrium. In the context of the early Universe we
have two reasons for departure from equilibrium. One is that if the
reheat temperature after inflation has been $T_{reh}$ then all
processes requiring energies larger than $T$ are suppressed and 
such processes play no role in establishing dynamical equilibrium. 
Thus particles that interact via only such processes remain out of 
equilibrium. We do not have much more control on such entities 
and in any case they most likely got inflated away and will not be
recreated due to insufficient energy for them to be created.

The more important source of departure from equilibrium is the
fact that the Universe is expanding. Like in an expanding gas, the 
temperature systematically falls. The primary assumption here is
adiabaticity -- i.e. extreme slowness of the rate of change
of temperature compared to the time scales of the equilibrating processes.
However, interesting epochs in the Universe correspond to times when
the rates of a few specific processes are becoming as small as the
expansion rate of the Universe. After the epoch is passed same reactions
go out of equilibrium and the last conditions remain impritned as
initial conditions for the rest of the evolution.

Schematically one can think of the Boltzmann equation as Liouville
operator $\hat L$ acting on distribution function $f$, with a driving force
provided by a collision operator $C$. In the absence of the collision
term we have equilibrium statistical mechanics.
$\hat{L} [f] = C[f]$
The Liouville operator which basically describes convection through the 
phase space can be written as
\[
\hat{L} = \f{d}{dt} + \vec{v} \cdot \vec{\nabla}_x + \f{\vec{F}}{m}
\cdot \vec{\nabla}_\nu
\]
assuming the conjugate momentum has the simple form of velocity times mass.
In General Relativity this has to be generalized to 
\[ p^\al \f{\p}{\p x^\al} - \Gamma^\al_{\beta\psi}
p^\beta p^\psi \f{\p}{\p p^\al}\]
which simplifies in the FRW case to
\[E \f{\p f}{\p t} - \f{\dot{R}}{R} \left|
\vec{p}\right|^2 \f{\p f}{\p E}
\]
The total number density is obtained by integrating the distribution
function over all momenta,
\[
n(t) = \f{g}{(2\pi )^3} {\dis \int} d^3 pf(E,t)
\]
Thus we obtain
\[
\f{g}{(2\pi )^3} {\dis \int} d^3 p \f{\p f}{\p t} - \f{\dot{R}}{R}
\f{g}{(2\pi )^3} {\dis \int} d^3p \f{\left| \vec{p}\right|^2}{E}
\f{\p f}{\p E} = \f{g}{(2\pi )^3} {\dis \int} C[f] \f{d^3p}{E}
\]
Exchanging the order of integration and differentiation in the first term
and working out the second term by doing an integration by parts, we
can show that this equation becomes
\[ 
\f{d}{dt} n + 3 \f{\dot{R}}{R} n = \f{g}{(2\pi )^3}
{\dis \int}
C[f] \f{d^3p}{E}
\]
Consider a process involving several particle species $\psi$, $a$, $b$...
\[
\psi + a+b \leftrightarrow i+j+...
\]
Our interest is usually a specific species which is undergoing an
important change. We think of the collision operator with species $\psi$ 
as the object of main interest to be
\bearrs
\f{g}{(2\pi )^3} C[f] \f{d^3p\psi}{dE\psi} &=& - {\dis
\int} \f{d^3 p_\psi}{(2\pi )^3 2E_\psi} \times \f{d^3p_a}{(2\pi
)^3 2E_a} \times... (2\pi )^4 \delta^4 \l(
p_4+p_a...-p_i-p_j...\r)\\[10pt]
&&\times \l[ \left| M\right|^2_\rar f_\psi f_a \l( 1 \pm f_a \r)
\right.\\
&&\left. - \left| M\right|^2_\leftarrow f_i \l( 1 \pm f_a \r) \l(
1 \pm f_\psi \r) \r] 
\eearrs
where $\mathcal{M}$ represents a matrix element for the concerned process.
There are Bose-Einstein and Fermi-Dirac distribution functions for the species 
in the in state. As for the out state, the $ \pm$ signs have to be chosen 
by knowing the species. Bosons prefer going into an occupied state 
(recall harmonic oscillation relation $a^\dag|n\rangle=\sqrt{n+1}|n\rangle$ 
so that an $n$-tuply occupied state has weightage proportional to $n$ to be
occupied). Hence the factors $(1+f)$, while fermions are forbidden from
transiting to a state already occupied hence the Pauli suppression factors
$(1-f)$. There is an integration over the phase space for each species.

Let us specialize the formalism further to cosmologically relevant case
with isentropic expansion. Since the entropy density $s$ scales as $1/R^3$, we
can remove the presence of $R^3$ factors in number density of non-relativistic  
particles by studying the evolution of their ratio with $s$. Thus we define 
the relative abundance for a given species $Y = \f{n\psi}{s}$. Then the we can
see that
\[
 \dot{n}_\psi + 3H n_\psi = s\dot{Y}
\]
Next the time evolution can be trade for temperature evolution in a
radiation dominated universe, by introducing a variable $x$.
\[
x \equiv \f{m}{T}
\mbox{so that}\quad t= 0.3 g_*^{-1/2} \f{m_{Pl}}{T^2} = 0.3 g_*^{-1/2}
 \f{m_{Pl}}{m^2} x^2 \equiv H^{-1}(m) x^2
 \]
Thus we get the equation
 \bearrs
 \f{dY}{dx}&=& - \f{x}{H(m)} {\dis \int} d\pi_\psi
 d\pi_a,...d\pi_i...(2\pi )^4 |M|^2 \delta^4 \l( p_{in}-p_{out}
 \r)\\[10pt]
 && \l[ f_a \; f_b...f_\psi - f_i \; f_j...\r]
\eearrs
Consider a species $\psi$ which is pair annihilating and going into
a lighter species $X$, $\psi\bar{\psi} \rar X\bar{X}$. The assumption is 
that the $X$ are strongly interacting either directly with each other or 
with rest of the contents, so that they equilibrate  quickly and remain in 
equilibrium. Thus the species to be studies carefully 
is $\psi$. Due to the property of the chemical potential and detailed balance 
which would exist if the $\psi$ are also inequilibrium  we can relate 
the equilibrium values
 \[
 n_Xn_{\bar{X}} = n^{eQ}_\psi n^{EQ}_{\bar{\psi}} = \l(
 n^{EQ}_\psi\r)^2 
 \]
Note the superscript $eq$ is not necessary in the $n_X$ due to it always
being in equilibrium.  We can thus obtain the equation
 \[
 \boxed{\f{dY}{dx} = - \f{x s}{H(m)} \left< \sigma_A | \nu |
 \right> \l( Y^2 - Y^2_{EQ} \r)}
 \]
The solution of above equations can be simplified by identifying convenient 
regimes of values of $x$ in which approximate analytic forms of $ Y_{EQ}$
exist
\bearrs
 Y_{EQ}(x)&=& \f{45}{2\pi^4} \l( \f{\pi}{8}\r)^{1/2} \f{g}{g_{*s}}
 x^{3/2} e^{-x} \;\;\;\;\;\; x \gg 3 \mbox{ non-relativistic case}\\[10pt]
 Y_{EQ}(x)&=& \f{45}{2\pi^4} \xi (3) \f{g_{eff}}{g_{*s}} = 0.278
 \f{g_{eff}}{g_{*s}} \;\;\; x \ll 3 \mbox{ relativistic case}
\eearrs
where the effective degeneracy factors $g_{eff}$ are defined relative to 
their usual values $g$ by $ g_{eff}= g_{boson}$ and $g_{eff}= \f{3}{4} g_{fermi}$.

\subsubsection{Freeze out and subsequent evolution}
We can get further insight into the special case considered above,
namely that of a species annihilating with its anti-particle and also going out 
of equilibrium.
Define
\[
 \Gamma_A \equiv n_{EQ} \left< \sigma_A | \nu | \right>
\]
which represents the rate of the reactions, given as a product of the microscopic 
cross-section $\sigma$, and number density times relative velocity as a
measure of the flux.
Using this, we can rewrite the evolution equation above in the form
\[
 \f{x}{Y_{EQ}} \f{dY}{dx} = - \f{\Gamma_A}{H} \l[ \l(
 \f{Y}{Y_{EQ}}\r)^2 -1 \r]\\[10pt]
\]
This shows that the rate of approach to equilibrium depends on two factors.
The second factor is the extent of departure from equilibrium, as we may expect
even in a laboratory process. The front factor $\Gamma_A/H$ represents 
the competition between the annihilation rate ( temperature dependent) and 
expansion rate ( also temperature dependent) of the Universe. 
When this front factor becomes small compared to unity, approach to equilibrium 
slows down, even if equilibrium is not reached. The abundance of the species 
$\psi$ in a comoving volume remains fixed once this factor becomes insignificant.
This phenomenon is called "freeze out", i.e., the fact that the relative abundance 
does not change after this and continues to evolve like free gas.

After the species freezes out, at epoch $t_D$ with corresponding temperature $T_D$, 
the distribution function of the species continues to evolve 
purely due to the effect of the expanding spacetime. There are two simple
rules of thumb we can prove for its distribution $d^3n/d^3p$ in phase space  :
\begin{itemize}
\item A relativistic species continues to have the usual Bose-Einstein or
Fermi-Dirac distribution function $(e^{\beta E}\pm 1)^{-1}$, except that
$\beta^{-1}=T$ scales like $T(t)=T(t_D)R(t_D)/R(t)$. 
\item A species which is non-relativistic, i.e., mass  $m\gg T_D$ the
number density simply keeps depleting as $R^{-3}$, just like the particles 
which are still in equilibrium. But the momenta scale as $R^{-1}$, so energy
$E=p^2/2m$ scales as $R^{-2}$. This is equivalent to the temperature scaling as
$T(t)=T(t_D)R^2(t_D)/R^2(t)$.
\end{itemize}

Thus the distribution functions have an energy dependence which is simply
obtained from their functional forms at the time of decoupling. In the relativistic 
case in fact remaining self-similar, and looks just like that of the particles
still in equilibrium, with an important exception. If there is a change
in the total number of effective degrees of freedom at some temperature, 
this information is  not conveyed to the decoupled particles. In the 
non-relativistic case the  scaling of the temperature parameter is significantly 
different. 

\subsection{Dark Matter}
There is a variety of evidence to suggest that a large part of the
matter content of the Universe is neither radiation, nor in the form 
of baryons. As such it is not capable of participating in processes
producing electromagnetic radiation and christened Dark Matter.

The direct evidence for Dark Matter is available at two largely 
different scales. At the scale of individual galaxies and at the scale 
of clusters of galaxies. At the level of single galaxies it is possible
to measure speeds of luminous bodies in the spiral arms for those
galaxies which are visible edge on. The difference in the redshifts
of the parts rotating away from us and the parts rotating towards us
is measurable. It turns out that as a function of their distance from the
center of the galaxy, velocities of rotation in the plane of the galaxy 
do not slow decrease in accordance 
with the $1/r^2$ law expected from Kepler's law. Rather their speeds 
remain steadily high even beyond the visible edge of the galaxy. 
The plots of the velocity vs. the radial distance from the center of
the galaxy have come to be called "rotation curves".
 The departure from Kepler law suggests presence of
gravitating matter extended to ten times the size of the visible galaxy!

Secondly at the level of clusters of galaxies, it is possible to measure
the relative speeds of the galaxies in a cluster, specifically the component 
of the velocity along the line of sight. By viirial theorem the values of 
these velocities should be set by the total matter content of the cluster. 
Again one finds the velocities more compatible with almost ten times the 
matter content compared to the visible.

Another indicator of the extent of the baryonic content is indirect but 
very sensitive.
Big Bang Nucleosynthesis predicts ratio of Hydrogen to Helium and the 
ratios of other light elements to Hydrogen determined by one parameter,
the baryon to photon ratio, $\eta=B/s$ where $B$ is the net baryon number 
(difference of baryon and antibaryon numbers) and the denominator is the 
photon entropy. We shall have occasion to discuss this in greater detail
in the section on Baryogenesis. The observed ratios of Helium to Hydrogen 
and other light nuclei to Hydrogen is correctly fitted only if $\eta\sim
10^{-9}$. Knowing the photon temperature very accurately we know the contribution 
of radiation to the general expansion ( it is very insignificant at present epoch).
Further knowing this accurately we know the baryon abundance rather accurately.
Between the two, the latter is certainly the dominant contribution to the 
energy density of the present Universe.
However the total amount of matter-energy required to explain the  current 
Hubble expansion is almost $30$ times more than the abundance baryons inferred
through the BBN data. Again we are led to assume the existence of other
forms of matter energy that account for the Hubble expansion. It is therefore
assumed that there is extensive amount of non-relativistic matter present
in the Universe today, and is called Dark Matter. We do not know at present
whether Dark Matter is a single species of particles or several different
species of particles. We do not know the possible values for the masses 
of such particles, however the study of galaxy formation suggests two classes 
of Dark Matter distinguished by their mass as we see in the next paragraphs.

The latest data from all 
sources suggest that the dominant component of the energy driving the expansion
is actually neither radiation nor matter, but some other form of energy
obeying an equation of state close to that of relativistic vacuum, $p= -\rho$.
This is estimated to be contribute about $70\%$. The Dark matter component
is estimated to be about $27\%$, and only about $3\%$ in the form of baryonic 
matter. These conclusions follow from WMAP data on CMB and type Ia supernova
data on expansion rate of the Universe more than 7 billion years ago.
It is remarkable that the approximately 10 times the abundance of Dark Matter
relative to baryonic matter as inferred directly from galacic and cluster data
is verified reasonably accurately by the very indirect methods. This is what
gives us confidence in the Dark Matter hypothesis.

When galaxy formation is considered this highly abundant Dark Matter component
plays a significant role. While no other kind of interaction is permitted between
baryonic matter and Dark Matter at least at low energies, gravity is always
a mediator. It is no surprise therefore that the Dark Matter is clustered
in approximately the same way as luminous baryonic matter. The question whether
there are large distributions of Dark Matter separately from baryonic matter 
needs  experimentally studied however so far the evidence does not seem to
demand such an assumption.

It then follows that the growth of perturbations which led to galaxy formation
must have proceeded simultaneously for the baryonic matter and the Dark Matter,
coupled to each other through gravity. The study of this coupled evolution
gives rise a distinction of two categories of Dark Matter which can be 
made based on the mass of the corresponding particle. Those particles that
have become non-relativistic by the time of galaxy formation are called
Cold Dark Matter ( CDM). They are in the form of pressureless dust by this epoch
and their chief contribution to energy density comes from their rest masses 
and not their thermal motion, hence Cold. We may think of this dividing line 
as set by the temperature $\sim 1 eV$ when neutral Hydrogen forms. Particles 
which are already non-relativistic at this temperature certainly belong to 
the category of CDM. 
On the other hand particles that remain a relativistic gas down to $1 eV$
temperature contribute through their thermal energy density and are called
Hot Dark Matter ( HDM). A prime candidate for this kind of DM is a neutrino,
whose masses are constrained to very small values. 

The main difference in the two kinds of DM comes from the nature of the 
clustering they assist. From Jeans formula we see that HDM clustering occurs 
at large physical scales while CDM can cluster at much smaller scales.
In fact too much HDM can destroy clumping of baryonic clusters at smaller
scales. Thus a study of the spectrum of perturbations $\mathcal{P}(k)$
gives a clue to the form of DM that assisted the formation of galaxies.
The current evidence in the light of the WMAP data strongly suggests
essentially the presence only of CDM, though some proportion of a HDM 
species cannot  be ruled out.

In the following subsections we shall show how we can trace back
at least some of the microscopic properties of the Dark Matter if we 
know its abundance today.

\subsubsection{Hot Relics}
For particles that continue to remain relativistic as they are going 
out of equilibrium, the equations from the previous subsection can
can be used to show that their abundance at late time is determined by
the value of their freeze out temperature, i.e.,  $x_{freeze\;out}$
\[
 Y_\infty = Y_{EQ} \l( x_{freeze\;out}\r) = 0.278
 \f{g_{eff}}{g_{*s}(x)}
 \]
If we want to think of this as the Dark Matter candidate, we estimate
the energy density it can contribute, which is determined to be
\[
 \rho_{\psi_o}= s_0 Y_\infty m = 3 Y_\infty \l( \f{m}{eV} \r)
\; \mbox{keV-(cm)}^{-3}
\]
From LSS data on distribution of fluctuations, as also the WMAP data 
it is now concluded that the structure formation could not have occurred
due to HDM. Hence this is not a very useful quantity to verify against 
observations.
Historically, this density value was used to put an upper bound
on the mass of a neutrino. If the decoupled neutrino is to not be 
so overabundant that it exceeds the current density of the Universe,
than its mass must be bounded.
 \[
 m \lesssim 13 eV \f{g_{*s}\l( x_f \r)}{g_{eff}}
 \]          
For $\nu$'s the ratio of the $g_*$ factors is $0.14$, from which 
one can conclude that $m_\nu < 91 eV$. This is known as the
Cowsik-McClelland bound. Although the bound is surpassed by both
by terrestrial experiments and recent astrophysical data, it is 
an instructive exercise.

\subsubsection{Cold Relics}
For cold relics,  we need to determine the quantities $x_f$, $T_f$ corresponding
to the freeze out of the species, and its present abundance relative to
radiation, $Y_\infty$. These are determined by solving the equation
\[
\f{dY}{dx}=-\f{1}{x^2}\sqrt{\f{\pi g_*(T)}{45}}M_P \langle \sigma v\rangle
(Y(T)^2 -Y_{eq}(T)^2 )
\]
It is useful to make an expansion of the cross-section in partial waves,
which amounts to an expansion in energy, or equivalently in the present 
setting, an expansion in $x=m/T$.  
For a massive particle the leading term is
 \[
 \left< \sigma_a | v | \right> \equiv \sigma_o \l( \f{T}{m}\r)^n =
 \sigma_o x^{-n} \; \; x \gtrsim 3
 \]
Thus expressing the cross-section as a function of $x$, the equation can 
be solved. The solution to this equation gives the left over abundance for 
a massive particle $\chi$ at present time. The answer typically has the 
following dependence 
\[
Y_\infty = O(1)\times\f{x_f}{m_\chi M_P \langle\sigma_A|v|\rangle}
\]
with $x_f$ determined numerically when the $Y$ effectively stops evolving.
The present contribution to the energy density due to these particles
is $m_\chi Y_\infty\times(s(T_0)/\rho_{crit})$ where $s(T_0)$ is the present 
value of entropy density in radiation.

It is thus possible to relate laboratory properties of the $\chi$ particle 
with a cosmological observable. Given a particle physics model, we can 
constrain the properties of the potential Dark Matter candidate by calculating
its contribution to $\Omega_{DM}$ and then counterchecking the same
cross-section in collider data.

\section{Missing Dark Energy}
The important topic of Dark Energy could not be included within the limitations of
this course. The reader can refer to some of the excellent reviews cited
at the end or await the next avatara of these notes.

\section{Baryon asymmetry of the Universe}

A very interesting interface of Particle Physics with cosmology is 
provided by the abundance of baryons in the Universe. At first
it is a puzzle to note that we only have baryonic matter present 
in the Universe, with no naturally occurring baryons to be seen.

In principle a cluster of galaxies completely isolated from others
could be made totally from anti-Hydrogen and anti-elements. However
there should be some boundary within which this confined, since
any contact with usual baryonic matter would generate violent
gamma ray production which would be observed as a part of cosmic
rays. But there are no clearly visible empty corridors separating 
some galaxies or clusters of galaxies from others, nor is there
a significant gamma ray background to indicate ongoing baron-anti-baryon
annihilation. Thus we assume the present Universe to be devoid of 
priordial anti-baryons. 

Due to charge neutrality, the electron number should be exactly equal
to the proton number of the Universe, and if Lepton number were conserved,
we should therefore have a predictable abundance of electron type 
anti-neutrinos. However after the discovery of neutrino oscillations
the question of total lepton number conservation is open and their
number may not be determined exactly by the charged lepton number.
Thus the total matter vs. anti-matter asymmetry of the Universe
is a more complicated question. We shall deal only with the baryons
where the situation is more precisely known.

The observed asymmetry is quantified by expressing it as a ratio
of the photon number density, i.e., entropy,
\[
\eta \equiv \f{n_B}{s} \equiv \f{n_b-n_{\bar b}}{n_\gamma}
\]
where the upper case subscript $B$ signifies the net baryon number 
while the lower case subscripts $b$, $\bar b$ signify the actual 
densities of baryonic and anti-baryonic species separately.
Big Bang nucleosynthesis constrains the value of this ratio very 
precisely. The abundances of Helium ${}^4$He to Hydrogen is sensitively
dependent on this ratio, but further, the abundances of light nuclei 
such as Deuterium, ${}^3\mbox{He}$, and ${}^7$Li relative to Hydrogen 
are also extremely sensitive to this ratio. 

\subsection{Genesis of the idea of Baryogenesis}
We believe that the Universe started with a hot Big Bang.
If the laws of nature were completely symmetric with respect to
matter and anti-matter both should be present in exactly same
abundance in thermodynamic equilibrium. Then the asymmetry between
the two has to be taken as an accidental initial condition.
Fortunately we know that the Weak interactions do not respect charge 
conjugation or matter-anti-matter symmetry $C$, but only the
product $CP$ after parity $P$ is included. Further, in 1964-65 two 
crucial discoveries were made.  It was shown that certain sectors 
of the theory ($K^0-\bar{K^0}$ system) also do not respect $CP$. 
In QFT there is a theorem that says that it is impossible
to write a Lorentz invariant local theory which does not respect
the combination $CPT$, now including time reversal $T$. Thus
a $CP$ violating theory presumably violates $T$ invariance in
same measure. The small mixing of $CP$ eigenstates will also be
reflected in small asymmetry in reaction rates involving these
participants.

The other crucial discovery was the Cosmic Microwave Background
which established the Hubble expansion back to temperatures as
high as $1000$K. It is easy to extrapolate this to sufficiently 
early times and higher temperatures when density and temperature
would be sufficiently high for Particle Physics processes to occur 
freely. The stage was set for searching for a dynamical explanation
for the baryon asymmetry (Weinberg 1964) and a specific model was 
proposed (Sakharov 1967). 

 \subsection{Sakharov criteria}
The minimal conditions required for obtaining baryon asymmetry
have come to be called Sakharov criteria. They can be understood
via a specific example. Consider a species $X$ which carries 
baryon number and is decaying into two different possible
final products, either two quarks or one anti-quark and a lepton. 
(We use one of the decay modes to determine 
the baryon number of $X$ and violation shows up in the other decay).
Such decays are easily possible in Grand Unified models. The
following should be true for a net $B$ number to remain in the
Universe :
 \begin{enumerate}
 \item Baryon number violation
 \[
 X \rar \barr{cc} qq&\Delta B_1 = 2/3\\
 \bar{q} \bar{\ell}&\Delta B_2 = -1/3
 \earr
 \]
 \item Charge conjugation violation
 \[
 {\cal M} (X \rar qq) \neq {\cal M} \l( \bar{X} \rar \bar{q}
 \bar{q} \r)
 \]
 \item CP violation reflected in difference in rates
 \[
 r_1 = \f{\Gamma_1(X \rar qq)}{\Gamma_1+\Gamma_2} \neq
 \f{\bar{\Gamma}_1 \l( \bar{X} \rar
 \bar{q}\bar{q}\r)}{\bar{\Gamma}_1 + \bar{\Gamma}_2} = \bar{r}_1
 \]
\item  Out-of-equilibrium conditions, which would make
reverse reactions become unfavorable
\bearrs
 \mbox{Net } B &=& \Delta B_1r_1 + \Delta B_2 \l( 1-r_1\r)\\[10pt]
 &&\l( - \Delta B_1 \r) \bar{r}_1 + \l( - \Delta B_2 \r) \l( 1 -
 \bar{r}_1 \r)\\[10pt]
 &=& \l( \Delta B_1 - \Delta B_2 \r) \l( r_1 - \bar{r}_1 \r)
\eearrs
 \end{enumerate}

In the early Universe, the condition for departure from Equilibrium
means that the reaction rate should become slow enough to be
slower than the Hubble expansion rate at that epoch. This will
happen because reaction cross-sections depend both on density
which is falling due to expansion, and the energy dependence 
of the intrinsic cross-section makes it smaller at lower temperature.
\bearrs
 \Gamma_X&\simeq& \al_X m^2_X/T\\[10pt]
H&\simeq&g^{1/2}_* T^2/M_{Pl} 
\eearrs
Need the rate $\Gamma_X$ still $< H$ when $kT \sim m_X$. Thus
$kT_D
\sim \l( \al_X \; m_{PL} m_X \r)^{1/2}$.\\

Resulting
\[
\f{n_B}{s} \simeq \f{B}{g_*} \times \mbox{ ( Boltzmann evolution)}
\]
Thus the result depends purely on the microscopic quantity $B$
(includes $\delta_{CP}$) and $g_*$ of the epoch when the mechanism
operates.

\subsection{Anomalous violation of $B+L$ number}

Quantization of interacting field theories contain many subtleties.
Sometimes the process of renormalization does not respect a symmetry
present at the classical level. Then quantum mechanically the
corresponding number operator is not conserved. This situation is
called anomaly. This is tolerable if it happens for a global charge.
If it happens for a gauge charge the model would not be viable.
It turns out that the Standard Model of Particle Physics does not 
respect the global quantum number $B+L$, baryon number plus
lepton number. The number $B-L$ is also not coupled to any gauge 
charge howeve it remains, miraculously, anomaly free and hence
is conserved.

The anomalous violation is not obtained in perturbation
theory. However  a handle on the anomalous violation rate can
be obtained by topological arguments involving global configurations
of gauge and Higgs fields. A specific configuration whose energy 
value determines the rate is called sphaleron. The energy of
a sphaleron is approximately 5 TeV in Standard Model. At temperatures
well below this, the rate is suppressed exponentially. At a
temperatrue much higher, the rate is order unity. Actually it 
becomes meaningless to speak of a conserved number. However
a number generated by any other mechanism will be quickly
equilibrated to zero by this non-conservation.

In the in between regime of temperatures, the rate is estimated as 
\[
\Gamma \approx \kappa \l( {\cal NV} \r)_0 T^4 \; e^{-E_{sph}(T)/kT}
\]
where $\kappa$ is the determinant of other fluctuations (recall
Coleman tunneling formula)  and ${\cal NV}_0$ represents sphaleron 
zero-mode volume, i.e., the weightage associated with all the 
possible ways a sphaleron can arise. 
This formula is valid for $m_W \ll T \ll m_W/\al_W$ where
$m_W$ is mass of the $W$ boson and $\al_W$ is the fine structure
constant $g^2/4\pi$ of the Weak interactions. 

Sphaleron energy depends on the Higgs mass at zero temperature
in such a way that too light a Higgs ($<90$GeV) would result 
in very rapid violation of $B+L$ around the electroweak phase 
transition. The  conclusion is that either the Higgs is heavier 
( which is corroborated by the bound $m_H>117$GeV from LEP data), 
or there is more than  one Higgs, or that there was a primordial 
$B-L$ number already present at the electroweak scale.

\subsection{Electroweak baryogenesis}
Could the baryon number arise at the elctroweak scale itself?
Sphaleronic processes are already becoming unimortant at this
scale. Also the properties any new particles needed can be 
counterchecked at the LHC or ILC. 
At the electroweak scale the expansion of the Universe is
many orders of magnitude ($10^{12}$) slower than the particle
physics processes. Hence direct competition with rates is
not possible. However, a first order phase transition leads
to formation of bubble walls. They sweep past any given point
in sufficiently short time scale that Particle Physics scales 
compete with  this time scale rather than the expansion time
scale of the Universe. Such a scenario which by-passes the
thermal conditions in the Universe is called \textit{non-thermal},
as against the example studied at the beginning of the
section which is called \textit{thermal} mechanism for baryogenesis.

If we enhance the SM with additional particles we can actually 
use the sphaleronic excitations to generate $B+L$ asymmetry
if the other criteria of Sakharov are satisfied.
Typical scenarios rely on 
\begin{enumerate}
\item Usual $C$ asymmetry of Weak interactions
\item $B+L$ violation by sphaleronic excitations
\item $CP$ violation due to complex phases in the vacuum 
expectation values of one or more scalar fields
\item Out-of-equilibrium conditions due a first order
phase transition
\end{enumerate}
It turns out that all of these conditions are easily satisfied
provided we have more than one Higgs scalar and sufficiently
large $CP$ phases entering some fermion masses. In specific
models favored for esthetic reasons however it has not been
easy to reconcile all the known constraints from other data
with the requirements of electroweal baryogenesis. For example,
the Minimal Supersymmetric Standard Model (MSSM) has the 
following dangers ( see M. Quiros hep-ph/0101230)
\begin{itemize}
\item Need for first order phase transition implies a light Higgs 
and a light superpartner "stop" of the top quark, as also a
bound on the ratio of the masses of the two neutral Higgs
bosons expressed as $\tan \beta$,
\[ 
110 < m_H < 115  GeV ,\; \;
\tan \beta \lesssim 4 \;, \;\; m_{\tilde{t}_R} \sim 105 \mbox{to} 165\mbox{GeV}
\]
\item One requires $\delta_{CP} \gtrsim 0.04$ which in turn raises 
the danger of color breaking vacua.
\end{itemize}

\subsection{Baryogenesis from Leptogenesis}
A realistic alternative possibility for dynamical explanation for
baryon asymmetry is thrown up by the discovery of neutrino mass.
The very small mass $m_\nu \sim 0.01$ eV for neutrinos requires their
Yukawa coupling to the Higgs to be $10^{-11}$. As we discussed in
case of inflation, such small dimensionless numbers seem to hide
some unknown dynamics going on. A very elegant explanation for
the small mass can be given if we assume (i) Majorana type masses
for the neutrinos and (ii) assume this mass, denoted $M_R$ to 
be high,  $M_R\sim 10^{14}$GeV. It can be shown that
\[
m_\nu M_R \simeq m_W^2
\]
is a natural relation if such a scale is assumed.
Now a scale like $10^{14}$ is also far removed from other physics, 
but is tantalisingly in the range of Grand Unified theories.  
This mechanism is called see-saw mechanism

This possibility makes leptogenesis naturally possible in the
early Universe almost literally by the example we studied earlier 
for the particle $X$ at the beginning of the section. Majorana
fermions do not conserve fermion  number. Further, the mixing
of the three generations can introduce a complex phase in 
the mass matrix which can lead to $CP$ violation. Finally
high mass means that the decay rate can actually compete
with the expansion scale of the Universe which is sufficiently
rapid at high temperatures, unlike at electroweak scale.
This can result in lepton asymmetry of the Universe.
This lepton asymmetry converts to baryon asymmetry as follows.
Recall that at temperatures high compared to
the electroweak scale, $B+L$ number is meaningless, and
will be equilibrated to zero. That is, the anomalous effects 
ensure $\Delta(B+L)=0$ and hence will generate approximately 
$\Delta B \sim -\Delta L$. The equality is not exact due to 
interplay of several chemical  potentials one has to keep 
track of.

An important handle on this proposal is provided by the
low energy neutrino data. It is possible to constrain the
extend of $CP$ violation that can be available at high 
scale from low scale masses due to see-saw mechanism.
Consider the decay of a heavy neutrino species $N$ into
a light lepton $\ell$ and a Higgs particle. There are several such
possibilities, and in each case the electric charge in the final state
is balanced to be zero. Due to lepton number violation characteristic
of Majorana fermions, the same $N$ can also decay into anti-lepton
and anti-Higgs. Thus the difference in the lepton number of the
final products in the two different modes is $\Delta L=2$ along 
the same lines as $\Delta B=1$ in our example at the beginning of the
section.
Then the $CP$-asymmetry parameter in the decay of any one
of the three  heavy neutrinos $N_i$, $i=1,2,3$ is defined as
\be
\epsilon_{i}\equiv\frac{\Gamma(N_i\rightarrow \bar{\ell}\phi)-
\Gamma(N_i\rightarrow \ell\phi^\dagger)}{\Gamma(N_{i}
\rightarrow \bar{\ell}\phi)+\Gamma(N_{i}\rightarrow \ell\phi^\dagger)}\,.
\label{epsilon_def}
\ee
If we assume a hierarchy of masses $M_1<M_2<M_3$ as is the case of
all other fermions, then the main contribution to the lepton asymmetry
generation comes from the species to decay last, i.e., the lightest
of the heavy neutrinos $N_1$. (Why?) 
The maximum value of CP violation parameter $\epsilon_{1}$ in this case 
can be shown to be
\be
|\epsilon_{1}|\leq 9.86\times 10^{-8}\left(\frac{M_1}{10^9\mbox{GeV}}
\right)\left(\frac{m_3}{0.05\mbox{eV}}\right)\,.
\label{epsilon1-max}
\ee
where the mass of the heaviest of the light neutrinos $\nu_3$ is
bounded by the atmospheric neutrino data, which gives
the mass-squared difference $\Delta m_{atm}^2\equiv m_3^2-m_1^2$. 
Thus, $m_3\simeq \sqrt{\Delta m_{atm}^2}=0.05$eV.

\begin{figure}[htbp]
\begin{center}
{\par\centering \resizebox*{0.5\textwidth}{!}
{\rotatebox{0}{\includegraphics{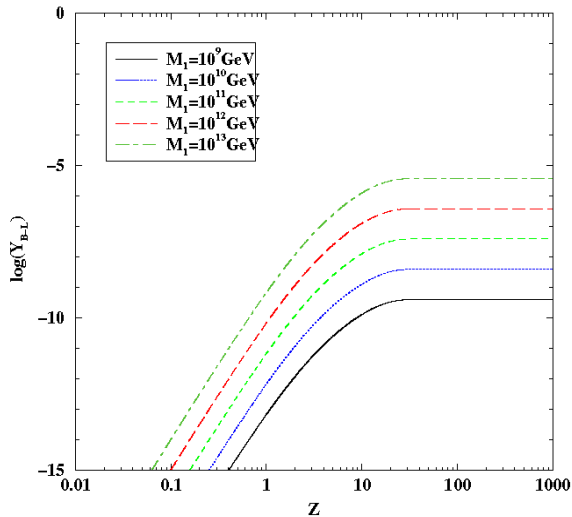}}} \par}
\caption{\small The evolution of the $B-L$ asymmetry with temperature,
shown here as a function of $Z=M_1/T$, with fixed values of $M_1$ as
indicated in the legend. The value of the $CP$ violation parameter is maximal
permissible and the parameter $\wt{m_1}$ explained in text is 
chosen $10^{-5}$eV for all graphs. Figure from N. Sahu et al,
Nucl. Phy. B752 (2006) 
}
\label{fig:Lgenesis}
\end{center}
\end{figure}

In the figure \ref{fig:Lgenesis} we show the solutions of the 
Boltzmann equations showing 
the accumulation of $B-L$ as temperature $T$ drops for various values
of $M_1$ with $CP$ violation chosen to be maximal permissible
according to above formula and the parameter $\wt{m_1}=(m_D^\dag m_D)_{11}/M_1$
chosen $10^{-5}$eV. It turns out that this particular parameter ( numerator 
is the $11$ element of the square of Dirac mass matrix for the neutrinos)
determines the overall outcome of the thermal $B-L$ number production.  
We see that there is negligible net number $B-L$ at high temperature
but it builds up as the decay processes are going out of equilibrium.
At some point the production levels off. Then due to sphalerons, the 
asymmetry which is initially  in the  form of light 
neutrinos also gets converted to baryonic form  producing net $B$ number.

From such exercises it can be shown that we need the mass 
scale $M_R$ of the heavy majorana neutrinos to be typically 
$>10^{12}$GeV but with some optimism, at least $>10^{9}$GeV for 
successful  thermal leptogenesis.
The problem with this conclusion is that firstly a 
new intermediate scale much lower than required for gauge
coupling unification is called for. Secondly, as discussed
in the Introduction, we expect supersymmetry to regulate
the QFT involved in Grand Unification with several scales
of symmetry breaking. But supersymmetry necessarily implies
the existence of gravitino. Further, it can be shown that
if our Universe underwent simple radiation dominated expansion
from any temperature larger than $10^9$GeV down to Big Bang 
Nucleosynthesis, sufficient number of gravitinos would be 
generated that would make the Universe matter dominated 
and foul up BBN. Thus it is usual to assume that the 
"reheat" temperature after inflation is lower than $10^9$GeV.
But then the thermal leptogenesis discussed here becomes
unviable.

It remains an open possibility that there are non-thermal 
mechanisms similar to the electroweak baryogenesis, but 
applicable to leptogenesis.

\section{Appendix}
\label{sec:app01}

Here we discuss the ``True or False'' statements given in 
section \ref{sec:GRrecap}. 
Note that some of the statements are half baked and warrant a
discussion rather than simple yes or no. Some hints.

\begin{enumerate}
\item Curved spacetime takes account of equivalence of gravitational
and innertial mass. The Relativity principle of space and time could 
have been Galilean and the formulation would be still useful.
See ref [1], chapter 12 for Cartan's formulation.

\item Reparametererization only relabels points. It cannot change
physics. Usually the laws are written only in the form invariant
under rigid rotations. But every law can in principle be rewritten
to be form invariant under change of parameterization. Thus 
reparameterization invariance cannot be a new physical principle.

\item Due to Equivalence Principle as adopted by Einstein, all 
forms of energy are subject to and contribute to gravitational field.
Energy density therefore must contain contribution of gravitational
"binding energy". However we can always choose freely falling frames 
locally so that effect of gravity disappear. In these frames the
energy density of gravitational field disappears. 

\item Total energy would be an integral of the energy density over 
a whole spacelike surface. This answer would remain unchanged 
under coordinate local transformations especially if we restrict 
ourselves to rigid transformations at infinity ( sitting where we 
measure up the energy). But GR throws up the possibility of
compact spacelike hypersurfaces. In this case asymptotic region 
is not available. 

\item If this genuinely means spacetime measurements are meaningless
at that point then it is unphysical. But it can be an artifact
of coordinate system, as for instance the origin in a spherical
or cylindrical coordinates.

\item Divergence of metric coefficients is often avoided using
different coordinate systems.

\item Curvature tensor is a physical quantity. Divergence of its
components will also often imply divergence of some components 
of energy-momentum tensor. Such points would be unphysical.
However note that much electrostatics is done assuming point
charges. These have infinite energy density at the location
of the point. When such points are isolated we  hope some 
other physics takes over as the singular point is approached.

\item The expansion of the Universe is neither relativistic,
nor a strong gravity phenomenon at least ever since BBN. It 
admits Newtonian description. If the spacelike hypersurfaces 
were compact that would be easier to explain as a dynamical 
fact in GR. In Newtonian physics we would simply accept is 
as fact, just as we are willing to accept infinite space as fact.
\end{enumerate}

\newpage
\section*{References}

\begin{enumerate}
\item 
Historical perspective along with personalities and an emphasis 
on a modern geometric  view of General Relativity can be found 
through out the textbook  
by C.~W. Misner, K.~S. Thorne and J.~A. Wheeler, \textsl{``Gravitation"} 
W. A. Freeman Co. (1973). 

\item
Emphasis on aspects of General Relativity which keep it on 
par with other field theories of High Energy Physics, is presented 
in S. Weinberg, \textsl{``Gravitation and Cosmology"} John Wiley and Sons (1973)

\item
  E.~W.~Kolb and M.~S.~Turner,
  \textsl{``The Early universe''}, Addison-Wesley Pub. Co. (1990)

\item T. Padmanabhan \textsl{"Theoretical Astrophysics, vol III : Galaxies
and Cosmology "} Cambridge University Press (2002)

\item  G.~Lazarides,
 \textsl{ ``Basics of inflationary cosmology''}, Corfu Summer Institute on Cosmology, 
  \textsl{J.\ Phys.\ Conf.\ Ser.}\  {\bf 53}, 528 (2006)
  [arXiv:hep-ph/0607032]. 

\item
A. Riotto \textsl{``Inflation and the Theory of Cosmological Perturbations"},
 Summer School on Astroparticle Physics and Cosmology, Trieste, (2002) arXiv:hep-ph/0210162

\item
R. H. Brandenberger \textsl{"Lectures on the Theory of Cosmological
Perturbations"} Lectures at summer school in Mexico, arXiv:hep-th/0306071

\item
T. Padmanabhan 
\textsl{``Understanding our Universe : current status and open issues"}
in \textsl{100 Years of Reality - Space-time Structure: Einstein and Beyond}, A. Ashtekar Ed., 
World Scientific Pub. Co., Singapore (2005), arXiv:gr-qc/0503107


\end{enumerate}

\end{document}